\documentclass[a4paper,fleqn,usenatbib]{mnras}
\usepackage{newtxtext,newtxmath}
\usepackage[T1]{fontenc}
\usepackage{ae,aecompl}
\usepackage{soul}

\usepackage{graphicx}   
\usepackage{amsmath}    
\usepackage{amssymb}    

\usepackage{hyperref}
\defcitealias{Schaye2015}{S15}

\newcommand{\msun}{M$_\odot$}
\newcommand{\mstar}{$M_{\star}$}
\newcommand{\GAR}{$\dot M_{\rm accr}$}
\newcommand{\simrec}{Recal-${\rm L25N752}$}
\newcommand{\simref}{Ref-${\rm L100N1504}$}
\newcommand{\units}{${\rm dex} \, r_{\rm 50}^{-1}$}

\title[Gas accretion and Radial Metallicity Profiles]{
The effect of gas accretion on the radial gas metallicity profile of simulated galaxies
}
\author[F. Collacchioni et al.]{
Florencia Collacchioni,$^{1,2}$\thanks{E-mail:fcollacchioni@fcaglp.unlp.edu.ar} 
Claudia D. P. Lagos,$^{3,4,5}$
Peter D. Mitchell,$^{6,7}$
\newauthor{
Joop Schaye,$^{6,}$ 
Emily Wisnioski,$^{4,8}$
Sof\'ia A. Cora,$^{1,2}$
and Camila A. Correa$^{6}$
}
\\
$^{1}$Instituto de Astrof\'isica de La Plata (CCT La Plata, CONICET,UNLP), Observatorio Astron\'omico, Paseo del Bosque,\\ B1900FWA, La Plata, Argentina.\\
$^{2}$Facultad de Ciencias Astron\'omicas y Geof\'{\i}sicas, Universidad Nacional de La Plata (UNLP), Observatorio Astron\'omico,\\ Paseo del Bosque, B1900FWA La Plata, Argentina.\\
$^{3}$International Centre for Radio Astronomy Research (ICRAR), M468, University of Western Australia, 35 Stirling Hwy,\\ Crawley, WA 6009, Australia.\\
$^{4}$Australian Research Council Centre of Excellence for All Sky Astrophysics in 3 Dimensions (ASTRO 3D).\\
$^{5}$Cosmic Dawn Center (DAWN), Niels Bohr Institute, University of Copenhagen$/$DTU-Space, Technical University of Denmark,\\ N{\o}rregade 10, 1165 K{\o}benhavn, Denmark.\\
$^{6}$Leiden Observatory, Leiden University, P.O. Box 9513, NL-2300 RA Leiden, The Netherlands.\\
$^{7}$Universit\'e Lyon, Universit\'e Lyon1, Ens de Lyon, CNRS, Centre de Recherche Astrophysique de Lyon UMR5574, F-69230,\\ Saint-Genis-Laval, France.\\
$^{8}$Research School of Astronomy and Astrophysics, Australian National University, Canberra, ACT 2611, Australia.
}
\date{Accepted XXX. Received YYY; in original form ZZZ}
\pubyear{2020}
\begin{document}
\label{firstpage}
\pagerange{\pageref{firstpage}--\pageref{lastpage}}
\maketitle

\begin{abstract}
We study the effect of the gas accretion rate (\GAR) on the radial gas metallicity profile (RMP) of galaxies using the {\sc eagle} cosmological hydrodynamic simulations, focusing on central galaxies of stellar mass \mstar$\gtrsim 10^9$~\msun\ at $z\le 1$. 
We find clear relations between \GAR\ and the slope of the RMP (measured within an effective radius), where higher \GAR\ are associated with more negative slopes. 
The slope of the RMPs depends more strongly on \GAR\ than on stellar mass, star formation rate or gas fraction, suggesting \GAR\ to be a more fundamental driver of the RMP slope of galaxies. 
We find that eliminating the dependence on stellar mass is essential for pinning down the properties that shape the slope of the RMP. 
Although \GAR\ is the main property modulating the slope of the RMP, we find that it causes other correlations that are more easily testable observationally: at fixed stellar mass, galaxies with more negative RMP slopes tend to have higher gas fractions and SFRs, while galaxies with lower gas fractions and SFRs tend to have flatter metallicity profiles within an effective radius.
\end{abstract}
%



\section{Introduction}
\label{sec:Intro}
%
The mass$-$metallicity relation (MZR) is the correlation between the gas-phase metallicity of galaxies and their stellar mass and has been established both in observations (e.g. \citealt{Tremonti2004, Maiolino2008, Troncoso2014}) and galaxy formation simulations (e.g. \citealt{DeRossi2015, DeRossi2017, Ma2016, Collacchioni2018, Tissera2019}) throughout a wide range in look-back time, $0$ to $\approx 12$~Gyr. 
The chemical enrichment of galaxies is expected to be the result of the interplay between inflows of pristine gas, outflows of enriched material and the star formation history of galaxies \citep{Finlator2008, Zahid2014a, Bothwell2016a}, and hence the study of the MZR holds important information about the formation of galaxies.

In the last decade, integral field spectroscopy (IFS) data such as CALIFA \citep{Sanchez2012}, KMOS \citep{Stott2014}, SAMI \citep{Bryant2015, Poetrodjojo2018} and MUSE \citep{Carton2018, Erroz2019} have greatly expanded the study of the MZR by providing spatially resolved information of the abundance of metals in galaxies, opening a new window onto the importance of chemical enrichment in the evolution of galaxies.\\
\indent
The study of radial metallicity profiles (RMP), i.e., the abundance of metals locked in the gas component of galaxies, typically measured through nebular emission lines (see \citealt{Maiolino2019} for a recent review), as a function of the distance to the centre of their galaxy, can help to understand how chemical enrichment takes place in galaxies. 
This is usually quantified as ${\rm log}_{10}(Z_{\rm gas}/Z_{\odot})={\alpha}\,r\,+\,b$, with $Z_{\rm gas}/Z_{\odot}$ being the gas metallicity in units of solar metallicity, and $\alpha$ being the slope of the RMP. 
Pioneering studies found that low-redshift galaxies ($z \leq 0.5$) display a negative RMP gradient ($\alpha<0$), i.e., galaxies are more chemically enriched in the central regions than in the outskirts \citep{Sanchez2012, Sanchez2013, Ho2015}. 
This is generally interpreted through the inside-out formation scenario, in which stars at the centre of galaxies form earlier than those in the outskirts and hence have had more time to enrich the interstellar medium (ISM), naturally producing higher gas metallicities in the centre compared to the outskirts \citep{Boissier1999}. 
However, further studies have found a variety of RMPs that complicates the picture. For example, \citet{Troncoso2014}, based on the AMAZE project, found that $3\lesssim z\lesssim 4$ galaxies display {\it positive} RMPs ($\alpha>0$), which the authors interpret as low-metallicity gas flowing directly into the centres of galaxies (see also \citealt{Cresci2010}).  

Instead of measuring a single $\alpha$ for the whole RMP, \cite{Sanchez2014} and \cite{SanchezMenguiano2016} using the CALIFA survey, and \citet{Belfiore2017} using the SSDS-IV MaNGA survey (among others authors), fit the RMP with two power laws, i.e. two different $\alpha$ for the inner and outer regions, finding puzzling results that cannot be easily interpreted within the inside-out scenario, such as a flattening in the outer regions, $\alpha \approx 0$, or relatively flat inner profiles (at least, less negative than expected) in high-mass galaxies ($M_\star \gtrsim 10^{10} \, {\rm M_\odot}$, \citealt{SanchezMenguiano2016}). 
This could be linked to feedback removing  metals preferentially from the inner regions of the galaxy (e.g. \citealt{Lagos2013,Muratov2017}), as well as significant star formation in the outer regions of the galaxy (for example, due to mergers or close interactions) that can quickly enrich the gas.

Several studies have explored the possible physical causes behind the shape of the RMP. 
According to \citet{Sanchez2012}, the change in the outer regions (beyond $\approx 2$ times the effective radius) could be due to different gas densities, the star formation history, or even the presence of bars which can alter the flow of gas internal to galaxies. 
Yet, \citet{Zinchenko2019}, also using the CALIFA survey, found no correlation between the RMP gradient and bars or spiral patterns, and \citet{Carton2018} did not find a correlation between $\alpha$ and the star formation rate (SFR) or stellar masses of galaxies when analysing MUSE data.
This is in tension with the reports of weak correlations between $\alpha$ and the specific SFR ($\rm sSFR ={\rm SFR}/M_{\star}$) by \citet{Stott2014} and \citet{Wuyts2016} using KMOS surveys, and with stellar mass by \cite{Ho2015} who use a combined CALIFA and SDSS Data Release 7 galaxy sample.
Moreover, \citet{Ho2015} found that measuring the RMP normalised by the galaxy's effective radius does lead to a correlation between $\alpha$ and galaxy properties, implying a co-evolution of gas and stars. 

It is fair to say that there is tension between the different observational results, which may in part be due to the different selection of galaxies, but also perhaps suggesting that the RMP is not well described by a single or double power-law, and/or that other galaxy properties may do correlate more strongly with the RMP than sSFR, stellar mass and morphology.
In fact, \citet{Carton2015}, measuring atomic hydrogen gas content (HI) with the WSRT of $50$ SDSS galaxies, found that the RMP slopes show a strong correlation with the HI mass fraction (the HI-to-stellar mass ratio), such that galaxies with higher HI mass fractions also have more negative $\alpha$.

From a theoretical perspective, a range of results from hydrodynamical simulations of galaxy formation have been reported that do not necessarily agree with each other. 
\citet{Tissera2016}, using a simulated cubic volume of $14$~Mpc on a side, argued that a range of slopes, from very negative to positive, can result from close galaxy encounters, presence of bars and/or low-metallicity gas accretion in the central regions of galaxies.
This variety of outcomes is caused by these processes triggering star formation activity and SN-driven outflows in different ways. 
A similar result is reported by \citet{Sillero2017}, who concluded that galaxy interactions generally lead to $\alpha>0$ for the case of major mergers of disc galaxies, but significant variations in the metallicity distribution and slopes are seen throughout the merger process. 
In addition, \citet{Sillero2017} found that the initial gas fraction of the galaxy merger and the SN feedback strength had an effect on $\alpha$. 
An important limitation of the studies above is the poor statistics, which makes it difficult to identify the primary processes that lead to higher/lower $\alpha$ values.
The latest generation of galaxy formation simulations has allowed a more thorough exploration of these trends with better statistics thanks to the much larger cosmological boxes available. 
One such example is the {\sc eagle} simulation (\citealt{Schaye2015,Crain2015}); its largest box is $100$~Mpc on a side, $\approx 360$ times larger volume than the studies above.
Using {\sc eagle}, \citet{Tissera2019} showed that very active merger histories are associated with galaxies displaying flatter RMPs, which naturally leads to lower mass galaxies having more negative RMPs.
\citet{Tissera2016} and \citet{Sillero2017} agree in that galaxies with stellar masses \mstar~$\leq 10^{10}$~\msun~show more negative RMPs than more massive ones. 
Other simulations, however, find contradictory results. 
\citet{Ma2017}, using cosmological zoom-in simulations from the FIRE project \citep{Hopkins2014}, reported low-mass galaxies to have flatter RMPs compared to massive galaxies. 
\citet{Tissera2019} also found no correlation between $\alpha$ and the environment in which galaxies live, while the physical size of the gaseous disc of galaxies appeared as one of the only galaxy properties clearly correlated with $\alpha$, in a way that more extended gas discs are associated with more negative $\alpha$.
Interestingly, \citet{Tissera2016} find that as redshift increases, high-mass galaxies (\mstar~$\geq 10^{10}$~\msun) show less negative and sometimes even positive $\alpha$. 

Gas accretion is expected to shape both the star formation and metallicity histories of galaxies and, hence, one would like to explore directly the correlation between the RMP and the gas accretion rate (\GAR). 
However, \GAR\ is not easily accessible in observations, motivating the simulation-based work to explore indirect tracers, such as star formation and stellar mass. 
Star formation is expected to be proportional to the gas supply, such that more infalling gas can trigger more star formation (e.g. \citealt{Dave2011}), while more massive galaxies live in more massive halos \citep{Behroozi2013a} and hence should be able to attract more gas into their gravitational potential. 
By studying directly the effect of gas supply onto galaxies, \citet{Perez2011} found that more negative slopes are associated with low-metallicity gas accretion, while \citet{Sillero2017} predicted a correlation between $\alpha$ and sSFR only in cases where the gas inflow triggers star formation within the central regions on short timescales. 
The proposed scenario is that significant \GAR\ onto the central regions of galaxies can increase the gas density, triggering high levels of star formation which rapidly enrich the inner regions of galaxies, leading to more negative slopes of the RMP. 
Other studies suggest extreme accretion, i.e. mergers, yields inverted metallicity gradients \citep{Troncoso2014}.
Hence, the study of RMPs might be essential to constrain the properties of the gas that is being accreted onto the galaxies (i.e. pristine or pre-enriched), and the effect \GAR\ has on their chemical evolution \citep{Finlator2017}.

The main limitation of the above studies is that they tend to focus on a small number of simulated galaxies, and hence it is difficult to separate effects related to gas accretion from other physical effects, such as galaxy mergers/interactions, depth of the gravitational potential well, and gas outflows. 
It is therefore essential to explore theoretically how \GAR\ affects the RMP of galaxies across cosmic time in a large sample of simulated galaxies over cosmologically representative volumes. 
This is the context in which our work develops. 
We use the {\sc eagle} cosmological hydrodynamical simulations suite to study how \GAR\ correlates with the RMPs, breaking the RMP into different regions of the galaxies. 
Our objective is to understand which regions of galaxies are most affected by the gas supply, what the roles of stellar mass, SFR and gas content are in determining the RMP, and what the ``smoking guns'' of gas accretion are that can be seen in the RMP. 
We do this by combining simulated boxes at different resolutions to allow us to cover the stellar mass range $M_{\star} > 10^9 \, \rm M_{\odot} $ at $0 \le z \le 1$.

This paper is organised as follows.
In Section~\ref{sec:model}, we briefly summarise the cosmological hydrodynamic simulations used in this work, as well as the galaxy sample selected, and introduce our definition of \GAR. 
We analyse how the RMP changes with \GAR, stellar mass, SFR and gas content in Section~\ref{sec:Results}, and show that \GAR\ is the property that is most strongly correlated with the slope of the RMP. 
In Section~\ref{sec:Interpretation}, we discuss the scope of our results and limitations, as well as the possibility of measuring the correlations found here in observations.
Finally, our conclusions are presented in Section~\ref{sec:Conclusions}.

\section{EAGLE simulations}
\label{sec:model}
The {\sc eagle} project\footnote{
http://icc.dur.ac.uk/Eagle/} 
is a set of cosmological hydrodynamical simulations that differ in box size, number of particles, numerical resolution and subgrid physics. 
A brief description of the simulations and the underlying subgrid models is presented in this section. 
For more information on the model, the reader can refer to \citet[][hereafter \citetalias{Schaye2015}]{Schaye2015} and \citet{Crain2015}, while the public data release is documented in \citet{McAlpine2016}.
To run the {\sc eagle} simulations, a modification of the N-body Tree-PM smoothed particle hydrodynamics (SPH) code \textsc{gadget}-3 \citep[last described by][]{Springel2005} is used, with the numerical methods  referred to as \textsc{anarchy} \citep{Schaller2015}.

{\sc eagle} follows the evolution of dark matter and gas particles, consistent with a flat $\Lambda$CDM cosmology characterised by the \citet{Planck2013} parameters: 
matter density $\Omega_{\rm m} = 0.307$, 
dark energy density $\Omega_\Lambda = 0.693$, 
baryon matter density $\Omega_{\rm b} = 0.04825$, 
square root of linear variance of matter distribution $\sigma_{\rm 8} = 0.8288$, 
index of power spectrum of primordial adiabatic perturbations $n_{\rm s} = 0.9611$, 
Hubble parameter $H_{\rm 0} = 100 \, h^{-1} \, {\rm km} \, {\rm s}^{-1} \, {\rm Mpc}^{-1}$ with $h = 0.6777$, 
and 
primordial helium abundance $Y = 0.248$.

A Friend-of-Friends \citep[\textsc{fof};][]{Davis1985} algorithm is used to identify haloes of dark matter (DM) particles; while the \textsc{subfind} algorithm \citep{Springel2001,Dolag2009} is used to identify the gravitationally bound subhaloes within a \textsc{fof} structure, linking the gas particles with their nearest DM particles. 
Theses subhaloes are then defined as the galaxies in the simulation. 
For each \textsc{fof} halo, the subhalo that contains the particle with the lowest gravitational potential value is defined as a central, and satellite galaxies are all of the remaining subhaloes.

{\sc eagle} contains a range of subgrid models that are described below, and those have free parameters that are calibrated to the observed $z \approx 0$ galaxy stellar mass function (GSMF), the relation between the galaxy stellar and black hole (BH) mass, and galaxy sizes of star forming galaxies. 
For this work, we use two of the {\sc eagle} simulations: 
the recalibrated high-resolution simulation of a $25^3$ cMpc$^3$ volume, which we will refer to as~\simrec, and the reference simulation of a $100^3$ cMpc$^3$ volume, called \simref. 
DM particle masses are $1.21 \times 10^6$~\msun~for the former and $9.7 \times 10^6$~\msun~for the latter, while the gas particle masses are $2.26 \times 10^5$~\msun~and $1.81 \times 10^6$~\msun, respectively.

The subgrid physical models implemented in the {\sc eagle} simulations are the following. 
Radiative cooling and photoheating on 11 elements (H, He, C, N, O, Ne, Mg, Si, S, Ca and Fe) are applied \citep{Wiersma2009a}, accounting not only for variations in metallicity but also in relative abundances. Hydrogen reionization is implemented as described in \citetalias{Schaye2015} at $z = 11.55$. 
Star formation follows the description in \citet{Schaye2008}, with a metallicity-dependent density threshold, in order to have a more realistic description of the gas transitions between warm, neutral and cold phases \citep{Schaye2004}. 
A Chabrier stellar mass function \citep[IMF,][]{Chabrier2003} is adapted, with minimum and maximum masses of $0.1$ and $100$~\msun, respectively. 
Stellar mass loss \citep{Wiersma2009b}, and mass and energy losses from Type II and Ia supernovae (SNe) are also implemented. 
Energy from SN is stochastically injected in nearby gas particles in the form of thermal feedback as in \citet{DallaVecchia2012}. 
In a similar way, feedback from active galactic nuclei (AGN) is computed with a fixed efficiency, and the energy is injected into the surrounding gas medium in a thermally and stochastic way (for details see \citetalias{Schaye2015}).

{\sc eagle}'s subgrid model imposes a gas temperature threshold that scales with the density of the gas (\citetalias{Schaye2015}) to prevent spurious fragmentation due to the finite resolution of the simulation.
Therefore, {\sc eagle} does not model the cold gas phase, which makes the interstellar medium of galaxies more homogeneous (or less clumpy) than we observe in real galaxies, leading to stars and dust in principle being more evenly mixed than it would be if a cold gas phase was included. 
Despite this limitation, as we increase the resolution of the simulation, though maintaining the temperature threshold above, the interstellar medium displays increased levels of clumpiness that is not numerically-driven and instead a physical effect \citep{Trayford2017,Trayford2020}. 
In {\sc eagle}, stars form in these clumps that have extensions of $\lesssim 1$~kpc, which is larger than the typical molecular clouds in the local Universe. 
As a result, we  expect metals to more easily mixed in the interstellar medium compared to what we would expect if more clumpiness was produced. 
A consequence of this could be less metallicity variations across the galaxy, but given we are interested in metallicity profiles measured in radial bins of $\approx 1$~kpc, this is likely not a big limitation. 
Current IFS observations used to constrain metallicity gradients have a similar spatial resolution to what we can achieve with the combination of {\sc eagle} simulations of \simrec\ for galaxies with \mstar~$< 10^{10}$~\msun\ and \simref\ for more massive galaxies ($500-1000$~pc for their median redshift; e.g. \citealt{vandeSande2019}). 

{\sc eagle} has proven to be very successful in a variety of comparisons.
Several papers have shown that the reference {\sc eagle} simulation reproduces observed galaxy properties such as the stellar mass function evolution from $z=0$ to $z=4$, the main sequence of galaxies \citep{Furlong2015}, the colour distribution of galaxies \citep{Trayford2015,Wright19}, and the gas content of galaxies of a given stellar mass \citep{Lagos2015, Lagos2016, Bahe2016, Crain2017}.
Very relevant for this work, the mass-metallicity relation of stars and gas, the resolved mass-metallicity relation \citep{Trayford2019}, as well as the gas metallicity gradients (\citetalias{Schaye2015}, \citealt{Tissera2019}), are well reproduced at \mstar~$\gtrsim 10^{10}$~\msun~for \simref. 
At lower stellar masses, deviations from the observations become important. 
The higher resolution, recalibrated run of the {\sc eagle} suite alleviates this problem, enabling agreement down to stellar masses of $10^9$~\msun~(see \citetalias{Schaye2015}, \citealt{DeRossi2017}). 
Hence, the simultaneous use of both the \simref\ and the \simrec\ simulations allows us to have very good statistics above $10^{10}$~\msun~in stellar mass (mostly from the \simref\ simulation), and to push down to $10^9$~\msun, thanks to the \simrec\ simulation, with confidence that the observed mass-metallicity relation is well reproduced over the entire stellar mass range. 
However, we keep the analysis of the two simulated samples separated, unless otherwise stated. 
We remind the reader that the \simrec\ simulation has slightly different parameters for the stellar and AGN feedback, as the aim of that run was to reproduce the $z=0.1$ stellar mass function with a higher resolution (see discussion on weak and strong convergence in \citetalias{Schaye2015}). 

The stronger feedback of the \simrec\ simulation tends to remove more metals from the ISM than what we see in \simref, which is partially why the metallicity of galaxies in the former is lower than in the latter at stellar masses $< 10^{10}$~\msun. 
\citetalias{Schaye2015} discussed this in some length and showed that a high resolution version of the \simref\ (with exactly the same subgrid physics model and parameters as the \simref\ run but with the volume and number of particles of the \simrec) produces gas metallicities that are $\sim 0.2$~dex higher than \simrec\ at stellar masses of $\approx 10^{8}$~\msun. 
Although the zero-point is slightly different between these two simulations, the slopes are overall in agreement once we studied them relative to their stellar mass (as we will see in the following sections).

\subsection{Computation of the gas accretion rate}
\label{sec:computation}
We compute \GAR\ onto galaxies using a simple particle tracking methodology \citep[following][]{Neistein2012}. 
For a given subhalo at a given simulation snapshot, we first identify gas particles that are classified as star forming (SF), as described in \citetalias{Schaye2015}. 
We then define accreting gas particles as the subset of these that are: 

\noindent
\textit{(a)} 
bound to the main progenitor subhalo at the previous (earlier) 
snapshot in the merger tree, and

\noindent
\textit{(b)} 
not star forming at that time. 

\noindent 
Star particles that fulfil conditions \textit{(a)} and \textit{(b)} are also taken into account in the calculation of the accreting gas. 
These stellar particles are those that come from gas particles that were accreted as of the previous snapshot. 
With this criteria, we ensure the accreting gas came from the same subhalo studied and, at the same time, that the accretion itself changes the state of the gas (from non-SF to SF).
In other words, we are interested in inflows that can trigger star formation. 
The latter is done to more easily connect with the observations, in which nebular emission lines are used as gas tracers and to measure the gas metallicity.

The choice of only counting the gas coming from the main progenitor is to compute the contribution of the accreted gas that comes from the diffuse circumgalactic medium (CGM) rather than galaxy mergers. 
We remark that previous hydrodynamical simulations show this ``smooth accretion'' to be the main source of gas accretion onto galaxies \citep{vandeVoort2011}. 
Wright et al. in preparation show that in {\sc eagle} $80$\% or more of the gas accreted onto halos comes from smooth accretion (which includes both pristine accretion and pre-processed gas). 
\GAR\ is then simply defined as the mass of accreted gas particles that satisfy conditions \textit{(a)} and \textit{(b)}, divided by the time interval between the 
snapshots. 
Progenitor subhaloes are identified using merger trees generated with the DHalo algorithm \citep{Jiang2014,Qu2017}. 

For the \simref\ simulation, we use $50$ snapshots to compute \GAR, while for \simrec\ we use $29$ snapshots. 
The reason why we use different numbers of snapshots, is that for \simref\ we have available a larger number of outputs with a better time cadence but with a reduced number of gas particle properties,
also called ``snipshots'', compared to the default $29$ snapshots that are publicly available from the {\sc eagle} database \citep{McAlpine2017}. 
For \simrec, however, we use the public data, which includes $29$ snapshots, with the complete list of particle properties. 
The time interval between outputs for \simref\ is $0.42$~Gyr from $z = 0 \rightarrow 0.03$, and $0.4$~Gyr from $z = 1 \rightarrow 1.05$, while for \simrec\ the time interval between outputs is $1.34$ Gyr from $z = 0 \rightarrow 0.1$, and $0.93$ Gyr from $z = 1 \rightarrow 1.26$. 
Although we are using different timesteps to compute accretion rates in the two simulations, this does not have an important impact on our results as accretion happens in long timescales (see Mitchell et al. in prep. and Wright et al. in prep.). 
If we were studying outflows, however, this difference in time cadence would be important, as outflows  happen in short timescales and are most stochastic. 
Hereafter, we use the word ``snapshots''  to refer to both these sets of outputs, making it clear which simulations we are referring to.

There is, however, an inherent weakness of the particle tracking method. 
If there is inflow that joins the SF ISM but leaves it before the snapshot in which the accreted gas is identified (with ``leave'' we mean that the particles are transferred to another phase, e.g. NON-SF ISM), then that inflow will not be counted. 
For more details about the method implemented here to measure the accreted gas, the reader can refer to Mitchell et al. (in preparation).

\subsection{Simulated galaxy sample}
\label{sec:sample}
Because of how sensitive the infalling gas is to environmental effects such as tidal stripping, ram pressure stripping, interactions, etc., which predominantly affect satellite galaxies, we limit this study to central, star-forming galaxies only (i.e. the galaxy sitting in the deepest part of the potential well of halos). 
We also apply the restriction of using only galaxies with \mstar~$\geq 10^9$~\msun~for the \simrec\ simulation and \mstar~$\geq 10^{10}$~\msun~for the \simref\ one, due to the resolution of the respective simulations (which gives us a minimum star particles number of $\approx 6700$ and $\approx 7700$, respectively). 
We include those central galaxies whose \GAR\ corresponds to at least $10$ SF gas particles, which translates to \GAR\ $\approx 9 \times 10^{-2}$~\msun~yr$^{-1}$ at $z = 0$ and \GAR\ $\approx 5 \times 10^{-2}$~\msun~yr$^{-1}$ at $z = 1$. 
In Appendix~\ref{sec:App-Resolution}, we test our main results against different choices of the latter minimum number of SF particles, and found them to be insensitive to thresholds up to $100$ particles. 
Our results are therefore well converged against this limit.

Our final sample comprises a total of $1,280$ galaxies at $z=0$ and $1,642$ at $z=1$. 
Because of the size evolution of galaxies (galaxies at higher redshift are smaller at fixed stellar mass, \citealt{Furlong2017}), we limit our study to $z \leq 1$, for which galaxies are better resolved. 
The median physical sizes of our selected galaxies\footnote{We compute the median value and dispersion of the half-stellar mass radius, $r_{\rm 50}$, for the whole sample at each redshift and simulation.} 
in the \simrec\ simulation are $3.98$~kpc and $3.17$~kpc at redshifts $z = 0$ and $1$, respectively. 
Similarly, the median galaxy sizes of our selected sample in the \simref\ simulation are $4.22$~kpc and $3.11$~kpc at redshifts $z = 0$ and $1$, respectively. 
Recall that the latter galaxies are more massive than the ones selected from the Recal-L25N752 simulation. 
In both cases, these sizes are comfortably above the gravitational softening of the corresponding simulation, which are $0.35$~physical kpc and $0.70$~physical kpc for \simrec\ and \simref, respectively. 
\citet{Ludlow2019d} showed that the spatial resolution of a simulation is approximately $0.05$ times the mean particle spacing, which for the \simref\ and \simrec\ are $66.5$~physical kpc and $33.35$~physical kpc at $z=0$, respectively, leading to a spatial resolution of $3.3$~kpc and $1.6$~kpc, respectively. 
This shows that the stellar mass selection applied here selects well resolved galaxies in these two simulations.

\subsection{Estimation of the gas metallicity profile}
\label{sec:metallicity}
Since we are interested in studying how the gas metallicity profiles are altered by \GAR, we need to define the way in which this profile is measured. 
Unless otherwise stated, we consider SF gas particles in spherical shells that increase in physical bins of $1$~kpc from $1$ to $9$~kpc, followed by bins of $3$~kpc width up to $51$~kpc, by bins of $10$~kpc until $101$~kpc, and by bins of $50$~kpc width up to $501$~kpc. 
This is done to avoid shot noise caused by the density of SF particles decreasing steeply with radius.
The metallicity of the smoothed gas \citep[\citetalias{Schaye2015},][]{Trayford2020} contained within each shell is defined as the ratio between the mass in metals heavier than helium over the total mass of gas, that is 
\begin{equation}
    Z = \frac{M_{\rm metals}}{M_{\rm H} + M_{\rm He} + M_{\rm metals}},
    \label{eq:GasMet}
\end{equation}
\noindent
where $M_{\rm H}$, $M_{\rm He}$ and $M_{\rm metals}$ are the masses of hydrogen, helium and metals of the SF gas particles, respectively. 
We use a logarithmic scale (${\rm log_{10}} (Z/Z_\odot)$) to express the metallicity at a certain shell, where $Z_\odot$ is the solar metallicity; we adopt $Z_\odot = 0.0127$ \citep{Asplund2005}.
This way of calculating the metallicity is common in theoretical works.
We note, however, that computing instead the oxygen to hydrogen abundance ratio (commonly used in observational works) gives effectively the same results.

\section{The effect of \GAR\ on the gas metallicity gradients}
\label{sec:Results}
In this section, we present correlations between several galaxy properties obtained by the {\sc eagle} simulations \simref\ and \simrec, i.e. stellar mass, \GAR, SFR and neutral gas fraction, and how they correlate with the RMPs. 
This is done at $z=0$ and $1$. 
The analysis has also been performed for redshifts in between, which produce consistent results. Hence, for the sake of conciseness, we only show $z=0,1$.

\subsection{The connection between metallicity, stellar mass and \GAR}
\label{sec:MZR}
\begin{figure}
  \includegraphics[trim=0mm 10mm 3mm 10mm, clip,width=1\columnwidth]
  {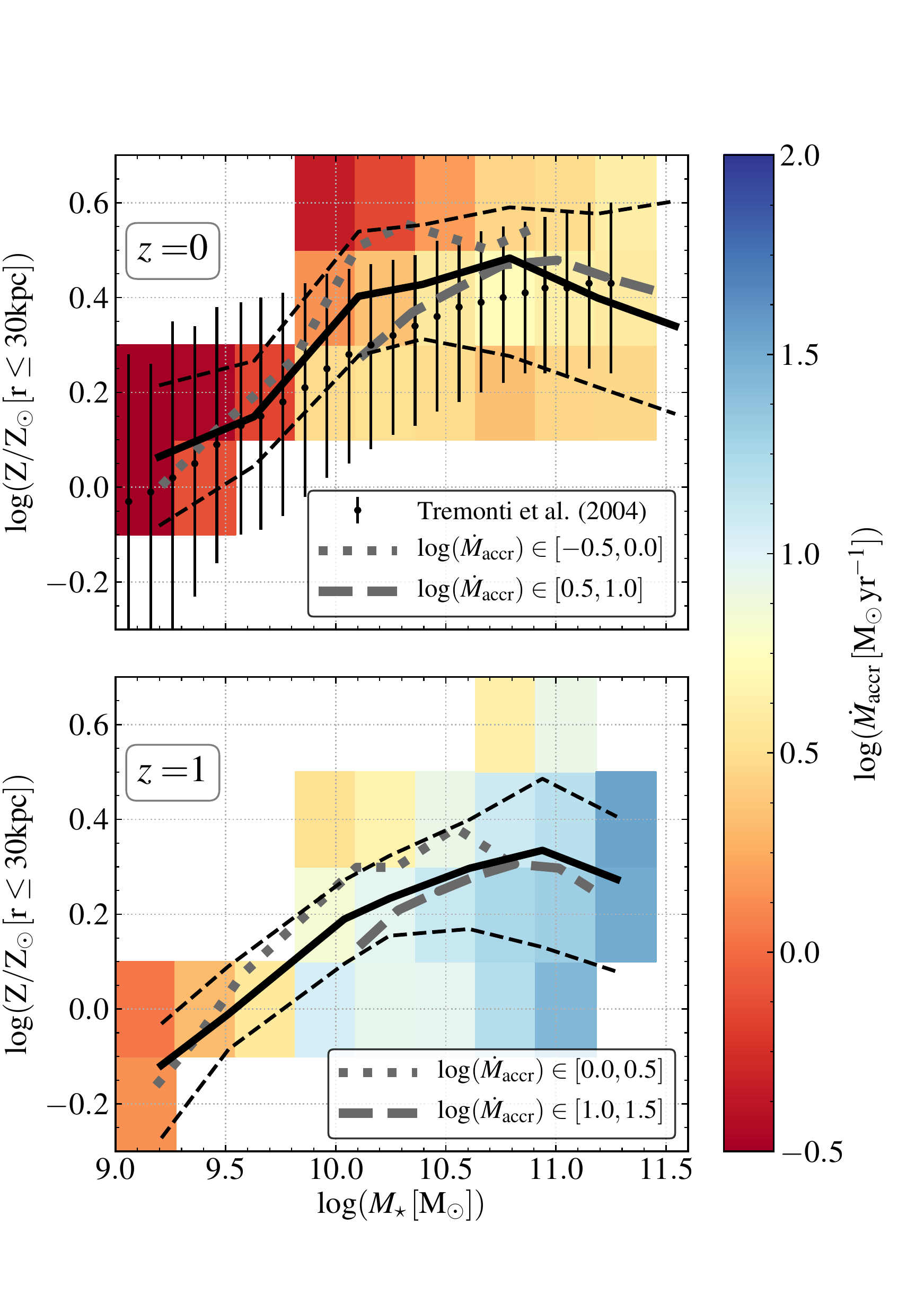}
  \caption{
  Stellar mass$-$gas metallicity relation (MZR) at $z=0$ (top panel) and $z=1$ (bottom panel) for galaxies in a combined simulated sample (combining the galaxy samples of the \simrec\ and \simref\ simulations; see details in Section \ref{sec:model}). 
  The gas metallicity is estimated using all the SF gas particles within $30$ kpc from the centre of potential of the galaxy. 
  In each panel, the black solid and thin dashed lines represent the median and $16^{\rm th}-84^{\rm th}$ percentiles, respectively. 
  Black symbols with errorbars in the top panel show the observational measurements of the MZR from \citet{Tremonti2004}.
  At fixed redshift, galaxies with higher stellar mass show higher gas metallicity.
  At fixed stellar mass, galaxies show a higher gas metallicity at lower redshift.
  Coloured pixels represent the median \GAR\ of the galaxies in bins of stellar mass and metallicity (containing at least $10$ galaxies).
  Galaxies with higher stellar mass show higher \GAR, with \GAR\ values decreasing as redshift decreases.
  Thick grey dotted and dashed lines show the median of galaxies in two bins of \GAR\ as labelled in the panels (presented in units of $\rm M_{\odot}\, yr^{-1}$).
  At fixed stellar mass and redshift, there is a tendency for the gas metallicity to be anti-correlated with \GAR.
  }
  \label{fig:MZR}
\end{figure}

Fig.~\ref{fig:MZR} shows the gas metallicity vs. stellar mass relation at redshift $z=0$ (top panel) and $z=1$ (bottom panel) for a combined sample of galaxies from the \simrec\ and \simref\ simulations, described in Section \ref{sec:model}.
In this case, the gas metallicity is estimated from all the SF gas particles within $30$~kpc from the centre of potential of the galaxy (this is the typical radius in which the integral properties of galaxies are calculated in the {\sc eagle} simulations; see \citetalias{Schaye2015} for details). 
Median values of the gas metallicity are shown as solid lines, while the dashed lines depict the $16^{\rm th}$ and $84^{\rm th}$ percentiles. 
Coloured pixels show the median \GAR\ of galaxies in bins of $0.25$ dex in stellar mass and metallicity. 
In observations, gas metallicities are usually estimated from regions that can be smaller than $30$~kpc. 
For this reason we investigated the effect of choosing a smaller aperture ($10$, $15$ and $20$~kpc) and find that changes in the MZR are negligible, except at the very high mass end, stellar masses $>10^{11} \, \rm M_{\odot}$, where we see an increase in the scatter. 
These differences, however, are small, and hence we decide to continue the analysis with an aperture of $30$~kpc to ease the comparison with other simulation papers.

As expected, the gas metallicity increases with stellar mass until a threshold mass above which the metallicity saturates or even decreases. 
This threshold mass increases with redshift from $\approx 10^{10}\rm \, M_{\odot}$ at $z=0$ to $10^{11}\,\rm M_{\odot}$ at $z=1$. 
This is related to the transition from galaxies that are preferentially star-forming to preferentially passive. 
At high redshift this happens at a higher stellar mass than at $z=0$, as seen from the development of the red sequence in {\sc eagle} \citep{Trayford2016,Wright19}. 
At $z=1$, \citet{Trayford2016} showed that the red sequence in {\sc eagle} is populated at the very high mass end (\mstar$>10^{11} \, \rm M_{\odot}$) by central galaxies and at low masses (\mstar$<10^{9.5} \, \rm M_{\odot}$) by satellite galaxies, with a dearth of galaxies at around $10^{10} \, \rm M_{\odot}$. 
This dearth region gets increasingly populated towards $z=0$. 
\citet{Trayford2016} and \citet{Bower2017} argued that this is due to AGN feedback being effective only at the highest masses at $z=1$, which is where black holes have grown enough to a regime where they can be efficient at driving outflows. 
The combination of the AGN effect and the lower number of star forming galaxies leads to a flattening of the MZR moving from higher to lower stellar masses as time goes on, and can even appear as a decrease of the metallicity at the high mass end. 
The latter has been reported in observations \citep{Sanchez2017}. 
This may in part may be due to systematic uncertainties. 
In these very metal-rich galaxies, it can happen that the electron temperature decreases, complicating the metallicity estimations \citep{Marino2013}. 
If we add to this scenario the fact that massive galaxies also have the higher \GAR\ at each redshift, then this inflow can be diluting the metallicity of the gas, contributing to the decrease seen in Fig.~\ref{fig:MZR}. 
At the low-mass end satellites only make a small contribution to the total number of galaxies and hence they do not impact the shape of the MZR.

The normalisation of the MZR also increases with decreasing redshift, and its shape (in terms of a slope of the relation) changes with redshift as well. 
Thus, the MZR evolves in normalisation and shape, implying that a galaxy's stellar mass and chemical abundance evolve at different rates. 
This has been reported by \citet{Guo2016} for the \simref\ simulation and by \citet{DeRossi2017} for \simrec. 
This evolution in normalisation is a natural success of the {\sc eagle} simulations and generally not easily obtained in other simulations (see \citealt{Collacchioni2018} for a discussion). 
It is interesting to note that, at fixed redshift, galaxies with higher stellar mass also have a higher \GAR, on average. 
This positive correlation is tight (about $\sim 0.5$ dex of dispersion) at all redshifts analysed in this work, and it is in agreement with previous theoretical works (e.g. \citealt{vandeVoort2011, Molla2016, Correa2018}, see also Appendix~\ref{sec:App-GARvsMstar}).

From Fig.~\ref{fig:MZR} it can also be seen that, at fixed stellar mass, the gas metallicity increases as \GAR\ decreases, regardless of the redshift considered.
Thick grey lines denote the MZR for galaxies in different ranges of \GAR\ (as it is described in the panels). 
Galaxies with lower values of \GAR\ have a MZR with a higher normalisation, i.e., with higher values of metallicity. 
This trend resembles the observed fundamental mass-metallicity relation plane \citep{Mannucci2010, LaraLopez2010}, which correlates the metallicity, the stellar mass and the SFR in the sense that at fixed stellar mass the metallicity increases as the SFR decreases. 
The existence of this relation in {\sc eagle} was demonstrated and discussed by \citet{DeRossi2017} and \citet{Matthee2018a}. 
In this scenario, the accretion of pristine gas has the dual effect of diluting the gas metallicity and triggering star formation \citep{Troncoso2014}. 
However, the fact that \citet{DeRossi2017} found a stronger anti-correlation between sSFR and metallicity in simulations with stronger stellar feedback indicates that outflows, which preferentially eject metals, also play an important role.

Galaxies have lower \GAR\ as redshift decreases (at fixed stellar mass), showing an evolution with time. 
This is expected as the overall gas accretion rate onto halos, and therefore onto galaxies, decreases with time \citep{Correa2018}. 
However, it is important to emphasise that large differences are expected between \GAR\ onto haloes and onto galaxies \citep[e.g.][]{vandeVoort2011, vandeVoort2017b, Nelson2013, Nelson2019}. 
The interplay between \GAR\ and SFR will be analysed in Section~\ref{sec:SFR}.

\subsection{The relation between the RMP and \GAR}
\label{sec:Maccreted}
\begin{figure*}
  \includegraphics[trim=4mm 0mm 0mm 0mm, clip, width=1.8\columnwidth]
  {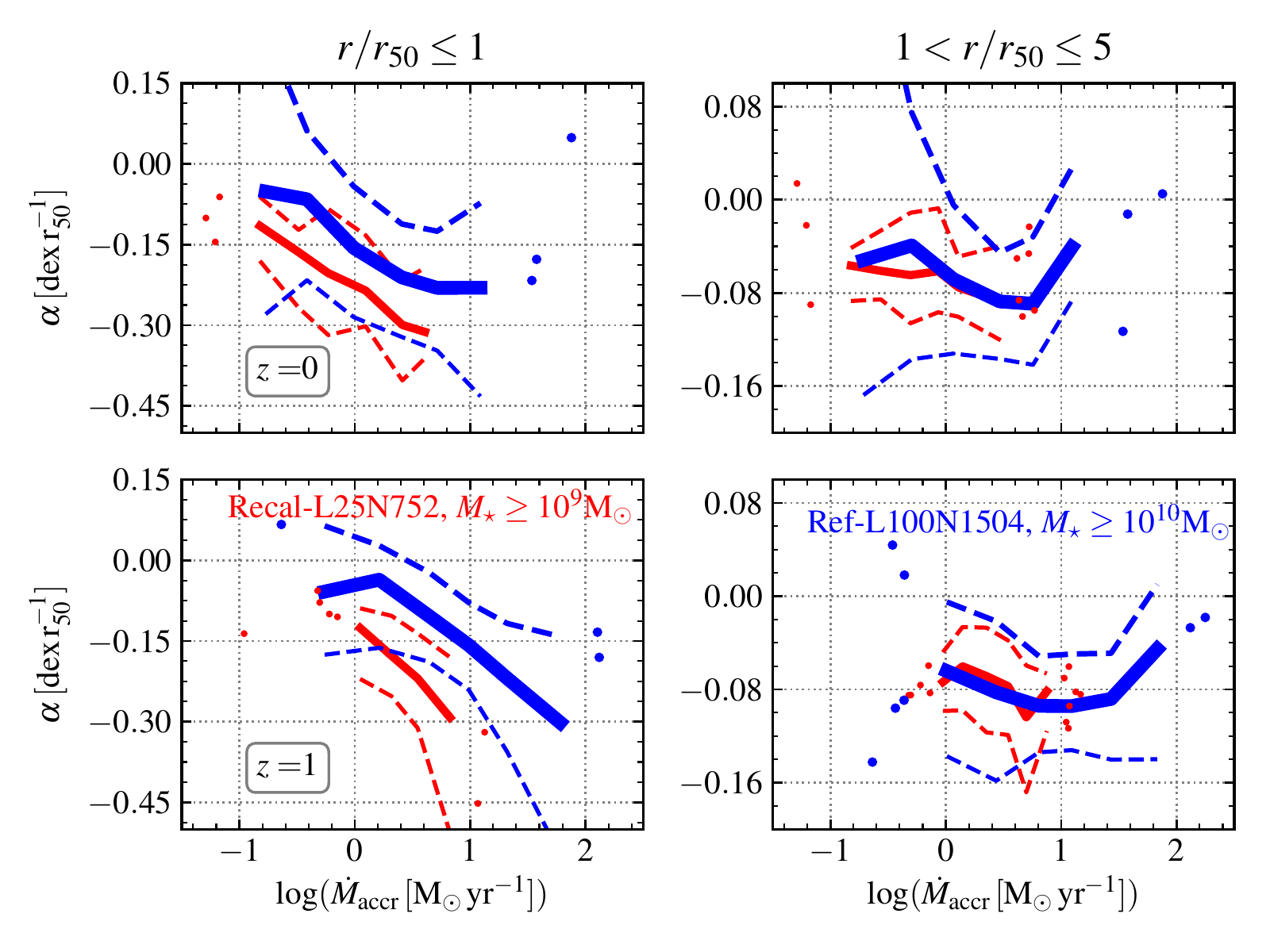}
  \caption{
  Median values of the RMP slopes, $\alpha$, for each simulation used in this work as a function of \GAR. 
  Top and bottom panels show galaxies at $z=0$ and $z=1$, respectively. 
  Left and right panels depict different regions of the galaxies, as labelled. 
  Galaxies are colour-coded by the simulation they are generated with, being red for \simrec\ and blue for \simref.
  Dashed lines depict the $16^{\rm th}-84^{\rm th}$ percentiles. 
  We show individual galaxies where bins have $<10$ objects (circles).
  The RMP slope in the inner region ($r/r_{\rm 50} \leq 1$) is negative and it is tightly correlated with \GAR\ for both simulations at all redshifts analysed, becoming more negative for higher \GAR. 
  The \simrec\ simulation shows a more negative slope, especially for higher values of \GAR. 
  At larger radii ($1 < r/r_{\rm 50} \leq 5$), we find a weaker correlation between $\alpha$ and \GAR\ with much larger scatter, with values of $\alpha \approx -0.08$. 
  As explained in Section~\ref{sec:sample}, the cuts in stellar mass for our galaxy samples are $M_\star \geq 10^9$~\msun~for \simrec\ and $M_\star \geq 10^{10}$~\msun~for \simref. 
  The difference in mass accounts for most of the difference between the simulations (see Section~\ref{sec:StellarMass}). 
  Note that the $y-$axis range changes from the left to the right panels, which is done to better highlight the values spanned by the data.
  }
  \label{fig:Slopes_Maccreted}
\end{figure*}
The aim of this work is to get a better understanding of how \GAR\ and other galaxy properties affect the RMPs.  
As was discussed in Section~\ref{sec:Intro}, the slope of the RMP measured over the whole galaxy does not depend on stellar mass or the environment in which the galaxy is immersed, particularly in the {\sc eagle} simulations \citep{Tissera2019}. 
However, since we are interested in finding out whether there is a preferential radius below/above which the profiles are significantly affected by gas accretion, we divide them into several regions and analyse the corresponding slopes instead of analysing a single slope of the RMP. 
We measure the power-law slope of the RMP, $\alpha$, in regions defined by the radii that contain half of the stellar mass of the galaxy, referred to as ${\rm r_{50}}$. 

The benefit of using the stellar effective radius is that we obtain $\alpha$ in regions that one would consider ``inner'' or ``outer'' regardless of the physical size of the galaxy. 
We measure $\alpha$ in two regions: $r<r_{50}$ and  $1<r/r_{50}<5$. 
We refer to these regions as inner and outer regions, respectively.
We do this by using a linear fit in the space of ${\rm log}(Z/Z_{\odot})$-$r$, as follows
\begin{equation}
    \log_{\rm 10}(Z/Z_{\odot}) = \alpha \times r + \log_{\rm 10}(Z_0/Z_{\odot}),
    \label{eq:Fitting}
\end{equation}
\noindent
where $Z$ is the gas metallicity defined by Eq.~\ref{eq:GasMet}, $\alpha$ is the slope we want to study (in units of \units, which allows us to study the change in the RMP slope with respect to a galaxy property independently of the physical size of  galaxies), and $Z_0$ is the normalisation of the fit. 
In this work we focus on $\alpha$ as the zero-point gas metallicity of galaxies has been studied in a variety of previous {\sc eagle} papers already (e.g. \citealt{Lagos2016,DeRossi2017}).

\citet{Tissera2019}, also using {\sc eagle}, found that the slope of the gas metallicity profile correlates strongly with the half-neutral gas mass size radius (i.e., the radius that contains half of the neutral, i.e. atomic plus molecular, gas mass); but that the half-stellar mass radius plays a small role. 
This suggests that the effective radius of the neutral gas content would be a more appropriate property to study the RMP. 
However, this property is only rarely available in observations and more difficult to define robustly. 
For example, observations of HI have shown that integrating longer, which allows to push towards lower HI column densities, continues to reveal HI, making the half-gas mass radius sensitivity-dependent \citep[e.g.][]{Oosterloo2007, Heald2011, Kamphuis2013}. 
The stellar $r_{\rm 50}$ is a well defined quantity that can be measured robustly for hundreds of thousands of galaxies (see \citealt{Lange2016} for an example in the nearby Universe).

Fig.~\ref{fig:Slopes_Maccreted} shows the relation between the slope of the RMP in the inner (left) and outer (right) regions of {\sc eagle} galaxies as a function of \GAR\ at $z \le 1$. 
Solid and dashed lines show the median values and $16^{\rm th}-84^{\rm th}$ percentiles, respectively. 
Results for both simulations are shown separately using different colours. 
There is a small offset between the two simulations in this figure, which is not surprising as for \simrec\ we are including all galaxies with stellar masses $M_{\star} \ge 10^9 \, \rm M_{\odot}$, while for the \simref\ simulation we limit the sample to galaxies with stellar masses $M_{\star} \ge 10^{10} \, \rm M_{\odot}$. 
We show later that eliminating the dependence on stellar mass brings the two simulations into agreement. 

For the inner region, it can be seen that galaxies with a higher \GAR\ display more negative RMP slopes $\alpha$, with $1\sigma$ scatter of $\approx 0.1$ \units.
The anti-correlation between $\alpha$ and \GAR\ is present in both simulations, although the trend is stronger for \simrec.
This trend is seen at all redshifts studied. 
A simple physical picture to interpret this trend would be the accreted gas having lower metallicity than the interstellar medium in the galaxy, and hence diluting the gas as it falls onto the disk. 
This would steepen the RMP because conservation of specific angular momentum will cause the accreted gas to settle in the outskirts of galaxies (see \citealt{Tissera2019} for an analysis of this scenario in {\sc eagle}). 
Another explanation is that the correlation in Fig.~\ref{fig:Slopes_Maccreted} may be caused by these two galaxy properties correlating with a third, more fundamental one. 
We explore this in upcoming sections.

In the outer regions, $1<r/r_{\rm 50}<5$, we find that $\alpha$ stays close to $\approx -0.08$~\units, with a scatter of about $\approx 0.25$~dex. 
It is interesting to note that the $\alpha$ values at $1<r/r_{\rm 50}<5$ are significantly less negative than the RMP slope in the inner parts for galaxies that have large \GAR, while galaxies with low \GAR\ have a similar slope at $r/r_{\rm 50}<1$ and $1<r/r_{\rm 50}<5$. 
There is a weak trend of \GAR\ being non-monotonically related to $\alpha$ in these outer regions, with $0.5 \leq \dot M_{\rm accr} {\rm [M_\odot \, yr^{-1}]) \leq 1.5}$ being associated to more negative values of the outer slope. 
This effect is however, weak, with differences of $\lesssim 0.05$~dex in $\alpha$ and only appreciable in the simulation \simref. 
The large scatter obtained in {\sc eagle} at $1<r/r_{\rm 50}<5$ is likely driven by the low numbers of SF gas particles galaxies have at their outskirts, making the measurement of $\alpha$ very noisy. 
We cannot therefore conclusively say that the relation between $\alpha$ and \GAR\ at $1<r/r_{\rm 50}<5$ is really non-monotonic, but we can assert that the outer parts of galaxies have only mildly negative RMPs. 
A possibility leading to the flat RMP at $r/r_{\rm 50} > 1$ is that the gas metallicity of the interstellar medium reaches that of the halo gas, which may act as a metallicity floor. 
To confirm this a more detailed analysis of the gas distribution should be made, which is beyond the scope of this work and will be left for future research. 
In addition, higher-resolution simulations would be required to increase the number of SF particles in the outer regions to better constrain the values of $\alpha$ there.

Previous simulation work have found galaxy mergers and close encounters to have a significant effect on their chemical enrichment patterns in a way that the inflowing gas from the interaction dilutes the metallicity of the central regions of the galaxy \citep{Montuori2010}. 
Observations of local ultra-luminous infrared galaxies suggest similar trends \citep{Rupke2008}. 
We explore this possibility by analysing the slopes of the galaxies that suffered a major or minor merger since the previous snapshot (i.e., galaxies that had more than one progenitor of $M_\star \gtrsim 10^8$~\msun). 
We find that only $\approx 10$ per cent of our sample has experienced a recent major or minor merger event, and that the slopes $\alpha$ in their inner and outer regions are very similar to the overall galaxy sample. 
Hence, we do not find any evidence that mergers affect the RMPs of galaxies in significant ways (at least at $z \le 1$). 
This in part may be due to the majority of galaxy mergers at $z \lesssim 0.5$ in {\sc eagle} having low gas fractions (neutral gas-to-stellar mass ratios $\lesssim  0.1$; see Fig.~$2$ in \citealt{Lagos2018b}) and hence having a reduced effect on the gas metallicity profiles. 
These are certainly different in nature to the objects studied in \citet{Rupke2008}, which were selected to be highly starbursting galaxies, and different to the simulations of \citet{Montuori2010} which are gas-rich major mergers. 
Therefore, it is not surprising that we find a weaker effect of galaxy mergers.

Given the strong (anti-)correlation we obtain between $\alpha$ at $r<r_{50}$ with \GAR, we analyse the latter in detail in the upcoming sections. 
From now on, unless otherwise specified, we only focus on the RMP at the inner region.

\subsection{The relation between the RMP and stellar mass}
\label{sec:StellarMass}
\begin{figure}
    \includegraphics[trim=3mm 4mm 0mm 3mm, clip,width=1.0\columnwidth]
    {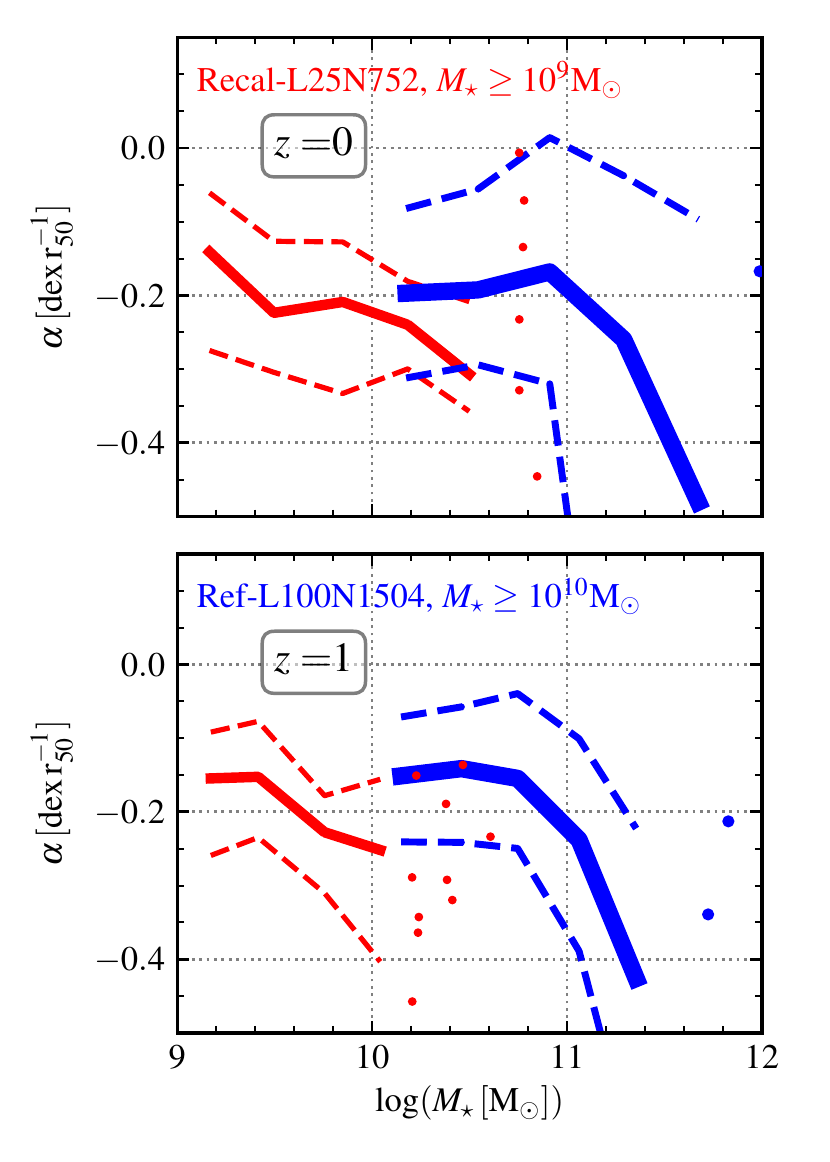}
    \caption{
    The median relation of the slope of the inner RMP, $\alpha$, as a function of the stellar mass, $M_\star$. 
    Top and bottom panels show galaxies at $z=0$ and $z=1$, respectively. 
    The \simrec\ simulation is shown with thin red lines, while the \simref\ simulation is shown with thick blue lines. 
    In all cases, dashed lines depict the $16^{\rm th}-84^{\rm th}$ percentiles, and individual galaxies are shown with circles where bins have $< 10$ objects.
    }
    \label{fig:Slope_Mstar}
\end{figure}
Fig.~\ref{fig:Slope_Mstar} shows the RMP inner slope, $\alpha$, as a function of stellar mass. 
Both simulations display an anti-correlation so that more negative $\alpha$ are associated with more massive galaxies. 
This may seem at first sight in contradiction to the results of \citet{Tissera2019}, who also used {\sc eagle}, but the main difference is that here we show $\alpha$ measured within $r_{\rm 50}$ while \citet{Tissera2019} measured $\alpha$ between $0.5 \times r_{\rm 50}$ and $2 \times r_{\rm 50}$. 
As a result, the RMPs in \citet{Tissera2019} are slightly flatter than the ones we calculate for $r<r_{\rm 50}$. 
To our knowledge there is no other result in the literature with the RMP slope calculated using the radial cuts that we adopt. 
As a consequence, we do not compare our results with observations and other theoretical findings due to the lack of consistency. 
We, however, encourage other works, specially observational ones, to use the radial cuts we propose here.

The two simulations analysed here, \simrec\ and \simref, in the stellar mass region when they overlap, are offset from each other, suggesting that stellar mass is not the primary property determining $\alpha$. 
This is not necessarily surprising, as the environments of galaxies with a stellar mass of $\approx 10^{10} \, \rm M_{\odot}$ in the \simrec\ are not necessarily the same as in the \simref\ simulation. 
\citet{Furlong2015} showed that the SFRs of galaxies of stellar masses $10^9-10^{10} \, \rm M_{\odot}$ are higher in \simrec\ than in \simref, while becoming more similar at $> 10^{10} \, \rm M_{\odot}$. 
This tells us that \GAR\ is higher in galaxies of stellar masses $\approx 10^{10} \, \rm M_{\odot}$ in \simrec\ than \simref\ (see Appendix~\ref{sec:App-GARvsMstar} for details regarding the \GAR$-$\mstar~relation).
We demonstrate below that \GAR\ is indeed the primary driver of $\alpha$ and that the offset seen in Fig.~\ref{fig:Slope_Mstar} is driven by the different \GAR\ \simref\ and \simrec\ galaxies of stellar masses $\approx 10^{10} \, \rm M_{\odot}$ have.

We quantify which property, \GAR\ or stellar mass, shows a stronger correlation with the inner slope $\alpha$, via the Spearman's rank-order correlation coefficient (${\rm R_s}$), finding that \GAR\ gives absolute values that are similar (\simrec) or larger (\simref, above $\approx 0.1$) than those obtained with stellar mass for all redshifts. 
Hence, in terms of scatter, $\alpha$ is better correlated with \GAR\ than with stellar mass.

\subsection{The relation between the RMP and \GAR\ at fixed mass}
\label{sec:GAR}
\begin{figure}
  \includegraphics[trim=3mm 4mm 0mm 3mm, clip,width=1.0\columnwidth]
  {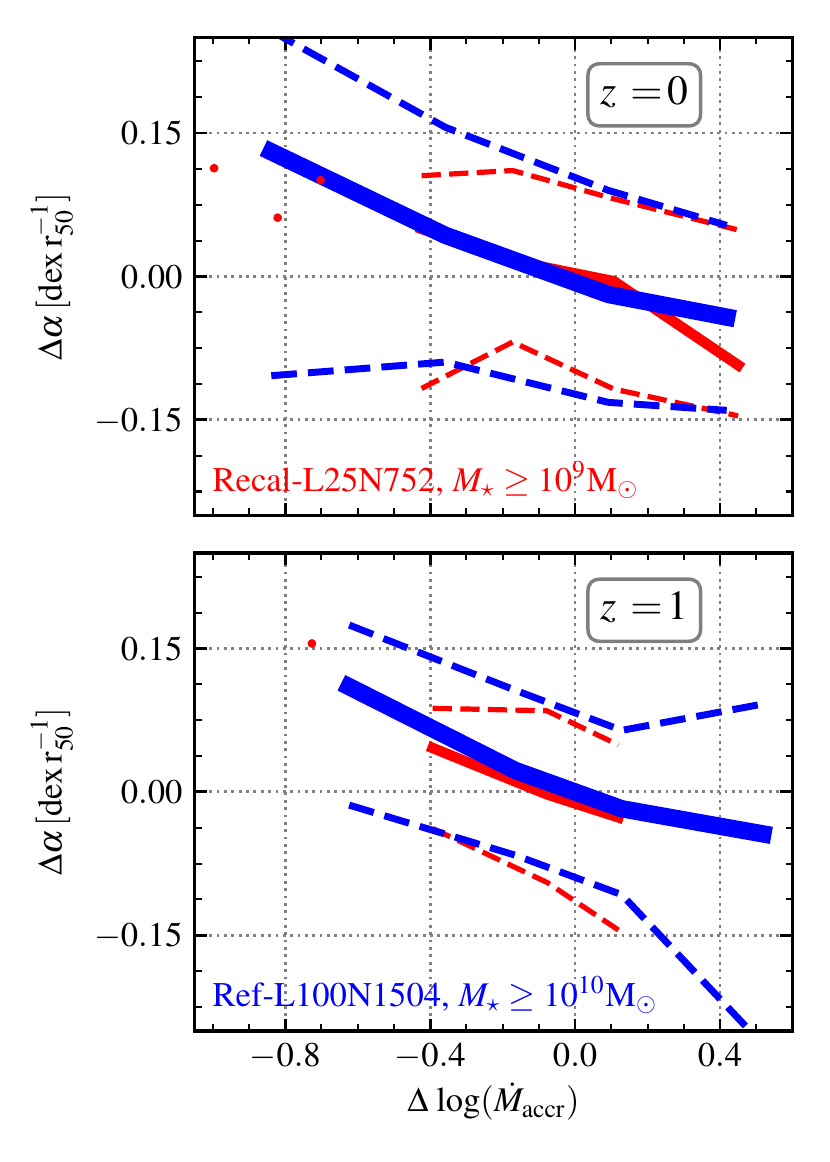}
  \caption{
  The residuals of the $\alpha$-$M_{\star}$ relation ($\Delta \alpha$) as a function of the residuals of the \GAR-$M_{\star}$ relation ($\Delta \log$(\GAR)), with $\alpha$ being the inner slope ($r/r_{\rm 50} \leq 1$) of the RMP. Top and bottom panels show galaxies at $z=0$ and $1$, respectively. 
  Red and blue represent the \simrec\ and \simref\ simulations, respectively.
  The $16^{\rm th}-84^{\rm th}$ percentiles are depicted as dashed lines, and solid lines show medians. 
  We show individual galaxies where bins have $<10$ objects (red symbols).
  Having eliminated the dependence on stellar mass, we still see a very strong anti-correlation between $\alpha$ and \GAR, indicating this to be independent of stellar mass. 
  The two simulations are in good agreement. 
  \simref\ covers a wider range of \GAR\ due to the larger cosmological volume compared to \simrec.
  }
  \label{fig:Residues_Maccr}
\end{figure}
We show in Fig.~\ref{fig:Residues_Maccr} the residuals of the $\alpha$-$M_{\star}$ relation ($\Delta \alpha$) as a function of the residuals of the \GAR-$M_{\star}$ relation ($\Delta \dot M_{\rm accr}$). 
We define these residuals as follows
\begin{equation}
    \Delta X = X \, - \, {\rm med}(X),
    \label{eq:residuals}
\end{equation}{}
\noindent
where $X$ is the property of interest of a galaxy, and ${\rm med}(X)$ is the median value at the stellar mass of the galaxy. 

By doing this calculation, we eliminate the stellar mass dependence from both axes. 
We find that the anti-correlation between $\Delta \alpha$ and $\Delta \dot M_{\rm accr}$ remains. 
We conclude that, at least for the inner regions, galaxies of fixed mass with a higher \GAR\ display steeper (negative) RMPs. 
Although there is a general expectation that higher \GAR\ can lead to changes in the RMP, to our knowledge this is the first time such results are quantified in cosmological simulations. 
Also note that for the residuals in Fig.~\ref{fig:Residues_Maccr} we do not see any differences between the standard and higher resolution {\sc eagle} simulations, suggesting that the differences seen in Fig.~\ref{fig:Slopes_Maccreted} were due to differences in their stellar mass distributions.

\subsection{The relation between the SFR, \GAR\ and the RMP}
\label{sec:SFR}
\begin{figure}
  \includegraphics[trim=3mm 4mm 0mm 3mm, clip,width=1.0\columnwidth]{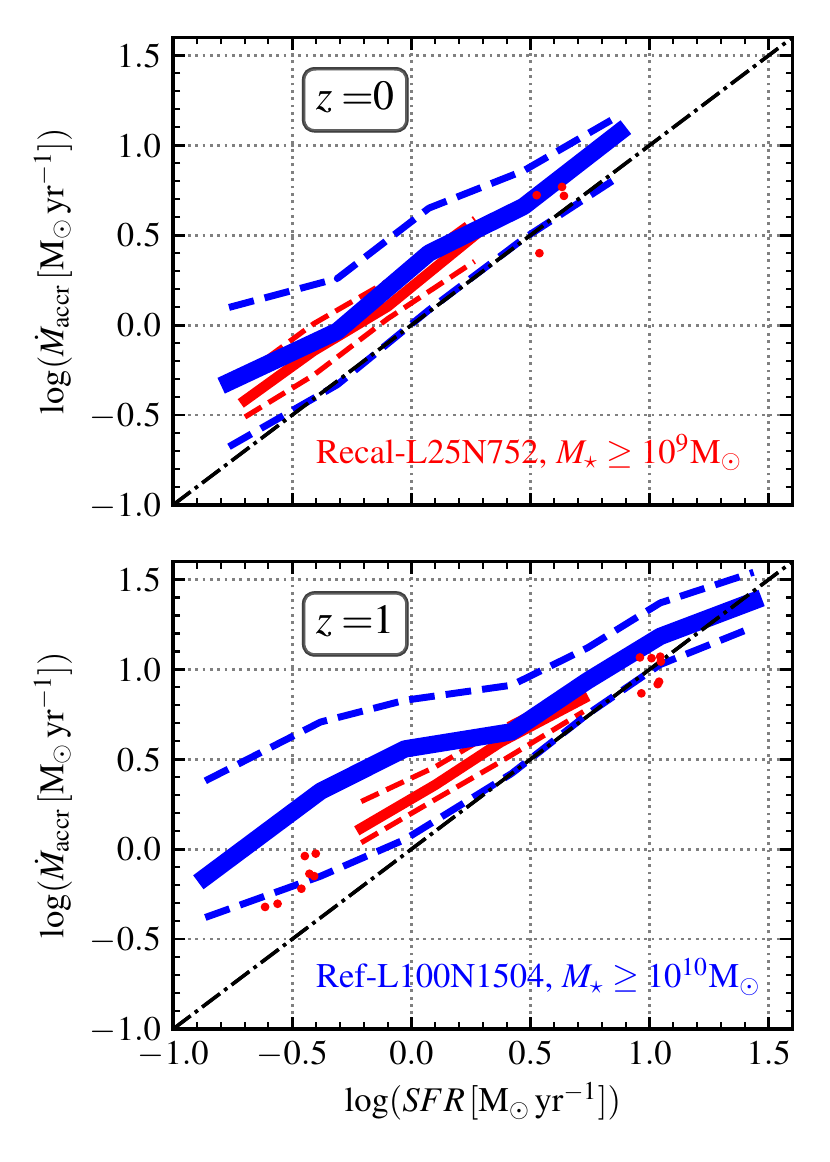}
  \caption{
  \GAR\ as a function of the SFR of galaxies in the \simrec\ (red) and \simref\ (blue) simulations. 
  Top and bottom panels show {\sc eagle} galaxies at $z=0$ and $1$, respectively. 
  The $16^{\rm th}-84^{\rm th}$ percentiles are shown as dashed lines. 
  We show individual galaxies in bins with $<10$ objects as symbols. 
  The dot-dashed lines show the 1:1 relation. 
  We remind the reader that \simrec\ galaxies were selected to have $M_\star \geq 10^9$~\msun, while \simref\ galaxies were selected to have $M_\star \geq 10^{10}$~\msun.
  Both simulations display a positive and tight correlation between \GAR\ and SFR. 
  Also, at fixed SFR, galaxies display a weak decrease in their \GAR\ of $\approx 0.2-0.3$~dex from $z=1$ to $0$ (only appreciable for galaxies with ${\rm SFR} < 10^{0.5} \, {\rm M_\odot yr}^{-1}$).
  }
  \label{fig:Maccreted_SFR}
\end{figure}
\begin{figure*}
  \includegraphics[trim=3mm 4mm 0mm 3mm, clip,width=1.8\columnwidth]
  {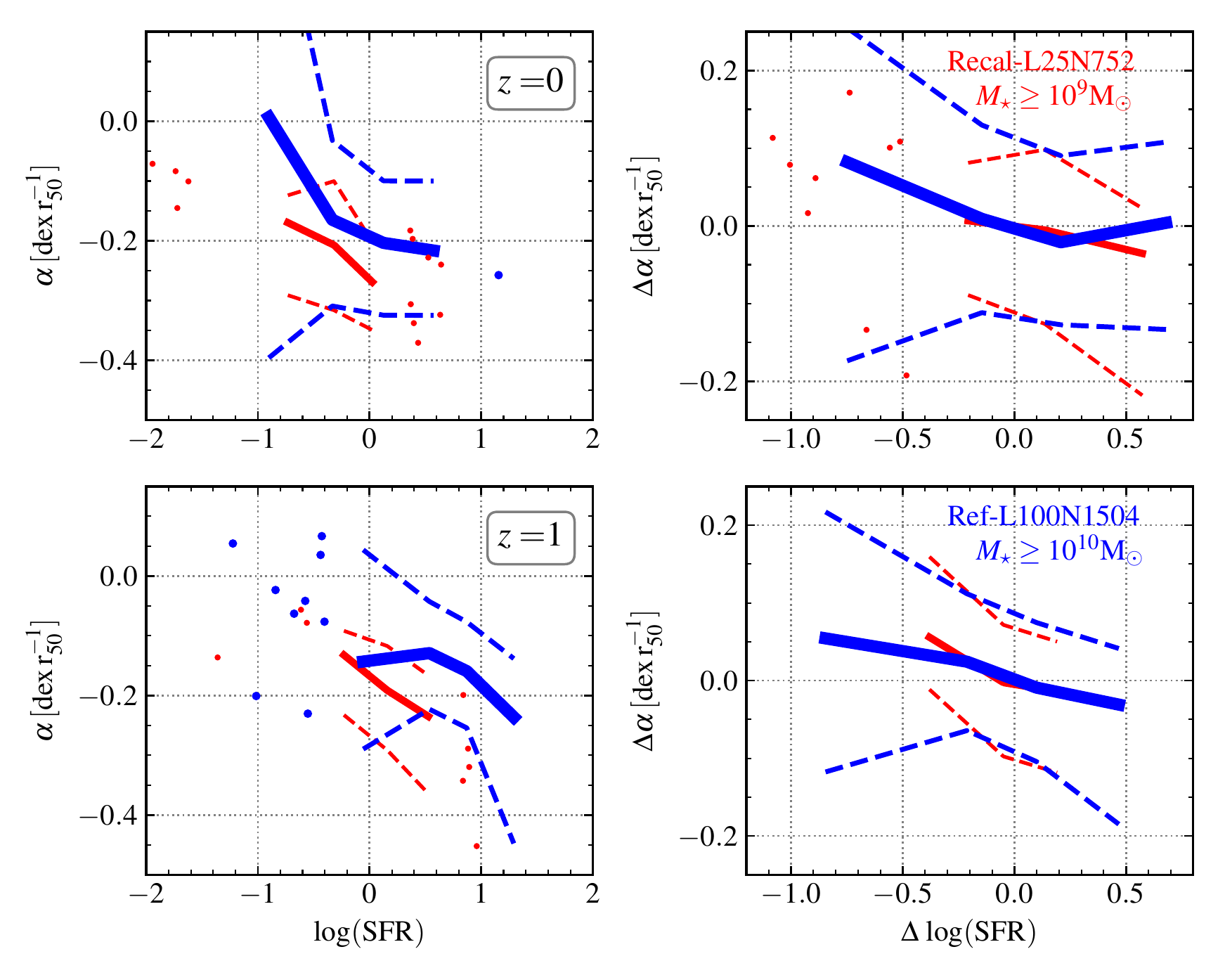}
  \caption{
  \textit{Left:} Solid lines show median values of inner ($r/r_{\rm 50} \leq 1$) RMP slope as a function of the SFR at redshift $z=0$ (top panel) and $z=1$ (bottom panel), for the \simref\ (blue) and \simrec\ (red) simulations.
  A similar anti-correlation is obtained as for $\alpha - \dot M_{\rm accr}$ (left panel of Fig.~\ref{fig:Slopes_Maccreted}). 
  \textit{Right:} Solid lines show median values of the residuals of the relation between the inner ($r/r_{\rm 50} \leq 1$) RMP slope and stellar mass, as a function of the residuals of the SFR and the stellar mass.
  Redshifts are as labelled. 
  Once we control for stellar mass (right panels), the anti-correlation is weaker than that between $\Delta \alpha$ and $\Delta \log(\dot M_{\rm accr})$ shown in Fig.~\ref{fig:Residues_Maccr}.
  This suggests that the anti-correlation between the slope of the RMP and the SFR is only a byproduct of the anti-correlation between the RMP slope and \GAR. 
  \textit{All panels:} 
  The $16^{\rm th}-84^{\rm th}$ percentiles are depicted with dashed lines.
  We show individual galaxies where bins have $<10$ objects (circles).
  }
  \label{fig:Slopes_SFR}
\end{figure*}
Observationally, measuring the gas accretion rate is very difficult as the accreted gas is expected to be low density and hence be very faint (e.g. \citealt{Fox2017}).
In addition, this inflowing gas is expected to have a relatively weak kinematic signature (compared to outflowing gas), which makes it challenging to separate spectroscopically from the intrinsic velocity dispersion of the resident ISM \citep{Bouche2017}.
On the other hand, the integrated SFR of a galaxy is much easier to infer and is expected to be closely correlated to the gas accretion rate. 
The latter is the basis for the ``equilibrium models'' of galaxy evolution. 
In these models, the gas accretion rate is perfectly balanced by the combination of SFR and outflow rates, and \GAR\ can then be inferred from the SFR and outflow rate. 
It follows then that \GAR\ is linearly correlated with the SFR (e.g. \citealt{Dave2012,Lilly2013}), modulated by the mass loading parameter (the ratio between the outflow rate and SFR).
As a consequence, one would expect the correlations discussed in Section~\ref{sec:Maccreted} to also extend to the SFRs of galaxies, i.e., that the RMP slope, $\alpha$, decreases with increasing SFR.

A correlation between $\alpha$ and the SFR of galaxies can have other interpretations beside the one provided by the ``equilibrium models''. 
For example, it is possible that the primary process controlling $\alpha$ is the metal enrichment of the ISM due to recently formed stars. 
In the latter case, one could imagine that the SFR may be the more fundamental parameter causing the change of the slope $\alpha$ rather than the gas accretion rate. 
This could be an interesting outcome since there still is debate as to whether there is a positive, negative or even null correlation between $\alpha$ and the SFR in galaxies (see Section~\ref{sec:Intro}).

Both of these two scenarios would result in an anti-correlation between $\alpha$ and the SFR, but in the former case one expects the scatter to be larger than for the relation between $\alpha$ and \GAR, while in the latter the opposite would be a more natural outcome.
To disentangle between these two cases, we study the same correlations of Fig.~\ref{fig:Slopes_Maccreted} and \ref{fig:Residues_Maccr}, but for the SFRs of galaxies. 

Before doing so, we verify the existence of a tight correlation between \GAR\ and SFR in Fig.~\ref{fig:Maccreted_SFR}. 
In agreement with the expectation of the equilibrium models described above, we find a strong correlation between \GAR\ and the SFR at all redshifts studied here, with a $1\sigma$ scatter $\lesssim 0.5$~dex. 
The scatter significantly decreases with SFR, from $\approx 0.5$~dex at $\rm SFR\lesssim 0.45\,\rm M_{\odot}\,yr^{-1}$ to $\approx 0.1$~dex at $\rm SFR \gtrsim 6 \, \rm M_{\odot} \, yr^{-1}$.
Although we do expect an overall correlation between the SFR and \GAR, the scatter here may be artificially small considering our definition of gas accretion, which is based on gas accretion that leads to star formation (see Section~\ref{sec:computation} and Mitchell et al. in prep. for details).
Interestingly, the predicted median relation evolves only weakly with redshift, at least at $z \le 1$, with differences of $\lesssim 0.3$~dex between $z=1$ and $0$ at fixed SFR.

The left panel of Fig.~\ref{fig:Slopes_SFR} shows the slope of the RMP in the inner regions ($r/r_{\rm 50} < 1$) as a function of the SFR. 
This relation shows a similar trend and scatter as the one with \GAR\ (see Fig.~\ref{fig:Slopes_Maccreted}). 
As we did in Section~\ref{sec:StellarMass}, we quantify the correlations, finding the $\alpha$-\GAR\ correlation to give a similar (\simrec) or higher (\simref) ${\rm R_s}$ by $\sim 0.1$ than the $\alpha$-SFR relation (again, \GAR\ reaching ${\rm R_s}$ values more negative than $-0.3$), suggesting that the more fundamental correlation is that with \GAR. 
\citet{Tissera2019} showed that {\sc eagle} galaxies with more negative slopes tend to have a larger fraction of their stars formed recently, consistent with the fact that there has been more gas accretion leading to such star formation activity, and with the clear anti-correlation we find here between $\alpha$ and the SFR. 
We caution, however, that \citet{Tissera2019} measured a single slope of the RMP rather than separating it into different regions of the disk.

As was done for \GAR, we remove the dependence on stellar mass by studying the residuals of the $\alpha$-$M_{\star}$ relation as a function of the residuals of the SFR-$M_{\star}$ relation (see Eq.~\ref{eq:residuals}) in the right panel of Fig.~\ref{fig:Slopes_SFR} for $z=0$ and $1$.
As explained in Section~\ref{sec:Maccreted}, the residuals are constructed as the difference between the property (i.e., slopes or SFR) and the median at the stellar mass of the galaxy. 
As was the case for Fig.~\ref{fig:Residues_Maccr}, controlling for stellar mass brings the two simulations into agreement. 
The simulation \simrec\ displays a weak anti-correlation between $\Delta \, \alpha$ and $\Delta \, \rm SFR$ (${\rm R_s}$, the Spearman's rank-order correlation coefficient, of $\approx -0.2$), similar to the anti-correlation between $\Delta \, \alpha$ and $\Delta \, \dot M_{\rm accr}$ (Fig.~\ref{fig:Residues_Maccr}). 
However, the simulation \simref\ shows a stronger anti-correlation with ${\rm R_s} \approx -0.3$. 
These results suggest that the $\alpha$-SFR relation is a byproduct of the other properties coming into play, such as that of \GAR\ and stellar mass. 
Therefore, we conclude that the inner RMP slope is primarily set by the gas accretion rate rather than by the SFR.

\subsection{The relation between the RMP and the gas fraction}
\label{sec:GasFraction}
\begin{figure*}
  \includegraphics[trim=3mm 4mm 0mm 3mm, clip,width=1.8\columnwidth]
  {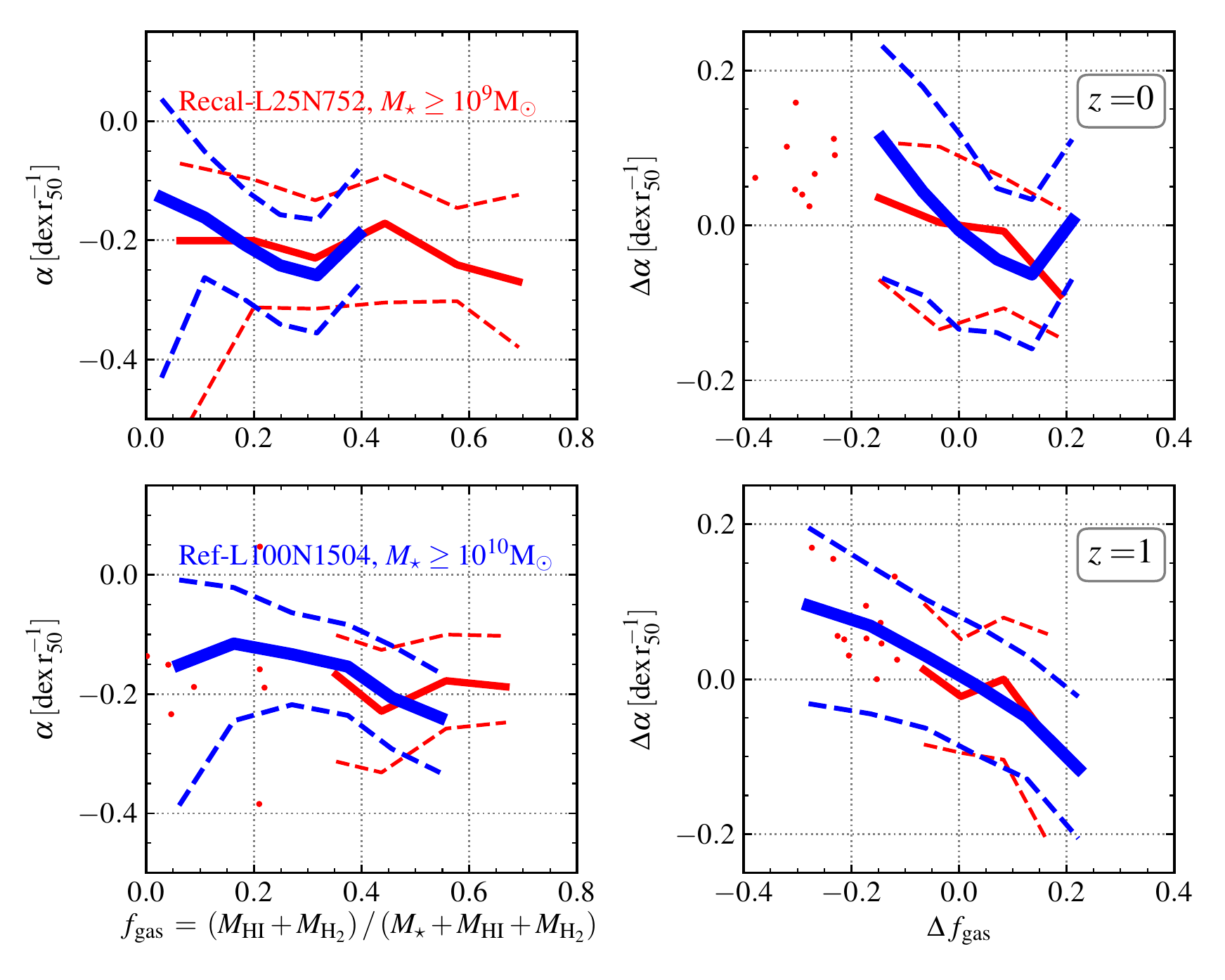}
  \caption{
  \textit{Left:} Solid curves show median values of the inner ($r/r_{\rm 50} \leq 1$) RMP slope as a function of the neutral gas fraction $f_{\rm gas}$ at redshift $z=0$ (top panel) and $z=1$ (bottom panel), for the \simref\ (blue) and \simrec\ (red) simulations.
  A similar anti-correlation is observed as in Fig.~\ref{fig:Slopes_Maccreted} for the inner part. 
  \textit{Right:} Solid lines show median values of the residuals of the relation between the inner ($r/r_{\rm 50} \leq 1$) RMP slope and stellar mass, as a function of the residuals of the $f_{\rm gas}$ and the stellar mass.
  Redshifts are as labelled. 
  As in Fig.~\ref{fig:Residues_Maccr}, the inner region RMP slope displays an anti-correlation with $f_{\rm gas}$ even after eliminating the dependence on stellar mass. 
  \textit{All panels:} 
  The $16^{\rm th}-84^{\rm th}$ percentiles are depicted with dashed lines.
  We show the individual galaxies where bins have $<10$ objects (red symbols).
  }
  \label{fig:Slopes_Fgas}
\end{figure*}
In the so-called ``equilibrium model'', \GAR\ regulates the gas content, SFR and metallicity of galaxies. The gas fraction of a galaxy is therefore expected to be a tracer of gas accretion (modulo the timescale to convert gas into stars and outflows). 
The gas fraction is attractive as is more readily available in observations than \GAR, and can be linked back to the latter (although under some assumptions). 
Hence, one would expect that a galaxy formation simulation which overall captures the nature of star-forming galaxies, produces a relation between a galaxy's gas metallicity and gas fraction. 
Here, we explore the relation between the RMP and the gas fraction of galaxies.\\
\indent
We use the neutral gas fraction measurement of \citet{Lagos2016}, defined as the neutral gas mass within $30$~kpc, $M_{\rm HI}+M_{\rm H_2}$, over the sum of the former and baryon mass in the same aperture, $M_{\star}+M_{\rm HI}+M_{\rm H_2}$. 
The neutral gas mass is found by applying the ionised-to-neutral gas separation in post-processing following \citet{Rahmati2013} (see \citealt{Lagos2015} for details of the subgrid phase partition of gas particles).\\
\indent
The left column of Fig.~\ref{fig:Slopes_Fgas} shows the RMP slope at $r<r_{\rm 50}$ as a function of the neutral gas fraction at $z=0$ and $1$ for the \simref\ (blue) and \simrec\ (red) simulations. 
If we first look at the \simref\ simulation, we see that there is a tendency for the RMP to decrease with radius more steeply as the gas fraction increases for all redshifts studied. 
This is because higher gas fractions are associated with higher \GAR. 
However, in the \simrec\ simulation we see a reversal of that relation. 
This is because the gas fraction is strongly anti-correlated with stellar mass (see Fig.~$1$ in \citealt{Lagos2016}). 
Hence, it is necessary to remove the stellar mass dependence to unveil a possible correlation between the RMP and gas fraction. 

The right column of Fig.~\ref{fig:Slopes_Fgas} shows the relation between the residuals of the RMP slope and the residuals of the gas fraction ($\Delta f_{\rm gas}$) as previously defined in Eq.~\ref{eq:residuals} (Section~\ref{sec:Maccreted}). 
The correlation is shown at $z=0$ (top panel) and $z=1$ (bottom panel). 
It is interesting to note that once the stellar mass dependence is removed, a strong anti-correlation is found, which is reminiscent of Fig.~\ref{fig:Residues_Maccr}. 
This could be related to the fact that all the galaxy properties mentioned in this work (i.e. \GAR, SFR and $f_{\rm gas}$) modulate, but to different degrees, the RMP slope at fixed stellar mass. 

We compute the Spearman's rank-order correlation coefficient ${\rm R_s}$ for the $\Delta \alpha - \Delta f_{\rm gas}$ relation (${\rm R_{s, gas}}$) and compare it with that obtained for the $\Delta \alpha - \Delta$~\GAR\ relation (${\rm R_{s, accr}}$). 
For both relations, ${\rm R_{s}}$ is similar for the simulations \simref\ and \simrec, with values of $\approx -0.2$ and $\approx -0.3$, respectively.
It is worth mentioning, though, that the absolute values of ${\rm R_s}$ indicate the $\alpha$-\GAR\ relation to be stronger than the $\alpha$-$f_{\rm gas}$ relation at all redshifts considered. 

It is curious that the relation of the RMP slope with  $f_{\rm gas}$ changes so much going from the left to the right panel of Fig.~\ref{fig:Slopes_Fgas}, compared to what happens with \GAR\ (Figs.~\ref{fig:Slopes_Maccreted} and \ref{fig:Residues_Maccr}) and the SFR (Fig.~\ref{fig:Slopes_SFR}). 
These large differences arise because the correlation between \GAR\ and gas mass (HI $+ {\rm H_2}$) changes with stellar mass. 
The relation is positive at low stellar mass, but flattens or even inverts as the stellar mass increases. 
Hence, at low stellar masses ($M_\star \lesssim 10^{10}$~\msun), gas accretion mostly increases the gas fraction of the galaxy (which is the case of dwarf galaxies), while at high stellar masses ($M_\star \gtrsim 10^{10.5}$~\msun), gas accretion may trigger AGN feedback which then reduces the gas fraction.
Thus, even though the relation between the RMP slope and $f_{\rm gas}$ gives us important constraints on the effect of gas accretion, it is neither as simple nor as direct a relation as the one we obtain between the RMP slope and \GAR.
\begin{figure*}
  \includegraphics[trim=4mm 0mm 0mm 3mm, clip,width=1.8\columnwidth]{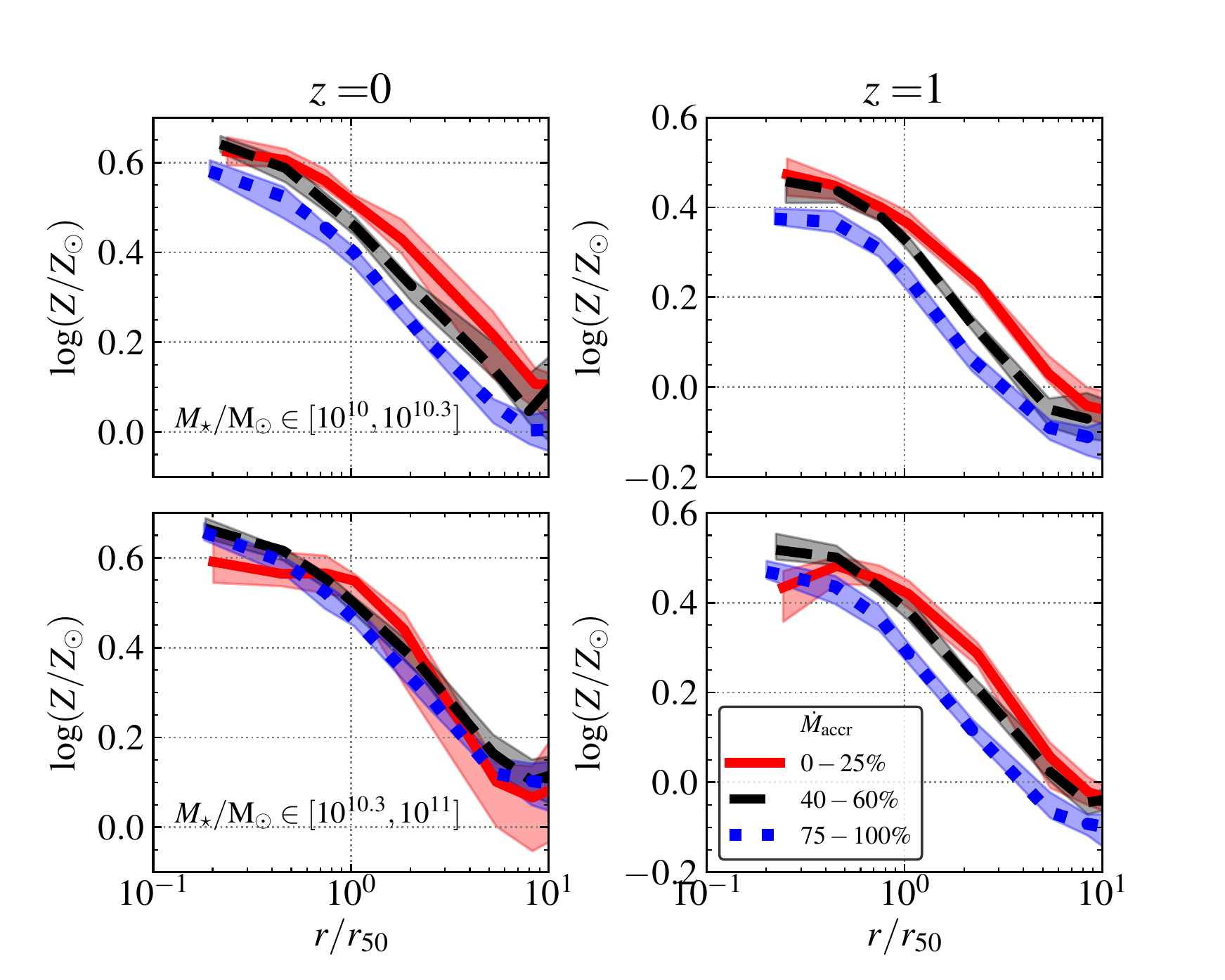}
  \caption{
  RMP at $z=0$ (left column) and $z=1$ (right column) for galaxies in the \simref\ simulation. The upper panels show galaxies with a stellar mass of $M_\star/M_{\odot} \in [10^{10}, 10^{10.3}]$, while the bottom panels show the range $M_\star/M_{\odot} \in [10^{10.3}, 10^{11}]$.
  Different colours represent the percentiles of \GAR, as labelled. 
  The lines and shaded regions show the median RMP and its $16^{\rm th}-84^{\rm th}$ percentiles, respectively.
  Galaxies with higher \GAR\ (dashed lines) show a steeper slope at $r<r_{\rm 50}$.
  Since the softening for the \simref\ simulation is $0.7$ kpc and we find average values of $r_{\rm 50} \approx 4.22$ kpc at $z=0$ and $r_{\rm 50} \approx 3.11$ kpc at $z=1$, the profiles will be resolved for $r/r_{\rm 50} \gtrsim 0.17$ and $r/r_{\rm 50} \gtrsim 0.23$, respectively.
  It is interesting to note that, at fixed stellar mass, there is a vertical offset of metallicity according to different values of \GAR, as was discussed in Section~\ref{sec:MZR}.}
  \label{fig:RMP}
\end{figure*}
See Appendix~\ref{sec:App-GasH2} for the results when only considering the molecular gas component (H$_2$) for the gas fraction, which can be of use since atomic gas (i.e, ${\rm HI}$) is currently not accessible in observations at $z \gtrsim 0.4$.

\section{Discussion}
\label{sec:Interpretation}
We showed in the previous sections (\ref{sec:Maccreted} and \ref{sec:SFR}) that the gas accretion plays an important role in shaping the inner slope of the RMP. 
We have also concluded that other galaxy properties, such as SFR, stellar mass and gas fraction, have a secondary role in the process of altering the slope, and are most likely driven by how these correlate with \GAR.
To better visualise the effect of gas accretion on the RMP, we show in Fig.~\ref{fig:RMP} the RMP of galaxies in bins of stellar mass and \GAR, for the \simref\ simulation, at $z=0$ (left) and $z=1$ (right).
The top panels show galaxies with stellar masses in the range of $M_\star/{\rm M}_\odot \in [10^{10}, 10^{10.3}]$, while bottom panels show a stellar mass range of $M_\star/{\rm M}_\odot \in [10^{10.3}, 10^{11}]$. 
We rank the \GAR\ values and present the median RMPs of galaxies in three \GAR\ percentiles: the bottom $25^{\rm th}$, between $40^{\rm th}-60^{\rm th}$, and the top $25^{\rm th}$. 
This allows us to compare galaxies in a way that is independent of the overall \GAR\ evolution.\\
\indent
Galaxies with higher \GAR\ show a steeper inner slope, independently of their stellar mass and redshift. 
In the stellar mass range $10^{10}-10^{10.3}\rm \,M_{\odot}$ at $z=0$ (top left panel in Fig.~\ref{fig:RMP}), galaxies have similar gas metallicities at the centre; however, galaxies with low \GAR\ have RMPs with a flattened core that can extend to quite large radii ($r \approx 0.5 \times r_{\rm 50}$), while galaxies with higher \GAR\ do not show a core in their RMP, and instead fall off sharply. 
This behaviour, however, is not universal (the top right panel does not show such a sharp fall). 
Note that at large radii, $r \approx 10 \times r_{\rm 50}$, galaxies display a flattening of their RMP, probably due to the ISM reaching the CGM metallicity. 
The latter happens at systematically smaller radii for galaxies with high \GAR.
In more massive galaxies (bottom left panel in Fig.~\ref{fig:RMP}), the flattened RMP core in galaxies of low \GAR\ is even more prominent, reaching out to $r \approx r_{\rm 50}$, while the higher \GAR\ galaxies do not display a core.
At $z=1$ the main difference is that the gas metallicities are overall lower than at $z=0$, and the differences in the normalisation of the RMPs at fixed stellar mass for different \GAR\ is larger. 

Curiously, profiles with the lower values of \GAR\ in massive galaxies (i.e., with stellar masses in the range  $M_\star / {\rm M_\odot} \in [10^{10.3}, 10^{11}]$) show a dip in the very internal parts of the RMP ($r < 0.5 \times r_{\rm 50}$). 
To understand this behaviour in more detail, we analyse the ratio between central black hole-to-stellar mass ratio ($M_{\rm BH}/M_\star$) of the galaxies in Fig.~\ref{fig:RMP}. 
We find that, independent of redshift, $M_{\rm BH} / M_\star$ decreases with increasing \GAR\ at fixed stellar mass, indicating that $M_{\rm BH}$ is the most massive for the lowest \GAR. 
In addition, we find that the most massive galaxies at $z=1$ of the bottom $25^{\rm th}$ percentile \GAR\ have a median $M_{\rm BH} / M_\star\approx 0.0007$, which is about twice the ratio of the other \GAR\ bins and stellar masses. At $z=0$ the trend between $M_{\rm BH} / M_\star$-\GAR\ is still present but with smaller differences between the different \GAR\ percentiles. This means that black holes in the subsample of the $z=1$ most massive, low \GAR\ galaxies have the most potential to affect their host galaxies.
This strong AGN activity is expected to remove the gas from the centre of the galaxy, specially the heavy elements, as the internal parts tend to be more metal-rich. 
Following the arguments in \citet{Bower2017}, gas in the centre of these galaxies forms stars and ignites a rapid BH growth phase that produces an outflow which expels significant amounts of gas and metals from the galaxy before these metals have time to mix with the surrounding gas. 
The dip seen in galaxies with \mstar/\msun$\in [10^{10.3}, 10^{11}]$ for the lowest \GAR\ at $z=1$ is therefore likely a product of the strong AGN feedback phase these galaxies are going through that removes metals from their centres before these metals have time to mix in the ISM.

Our analysis can also give insight into where the gas recently accreted onto the galaxy is. 
Galaxies with higher \GAR\ display steeper gradients in the internal parts. 
Thus, gas accretion appears to be diluting the metals in an outside-in fashion, which translates into more negative RMPs.

An important consideration to interpret the observed trends above is the way in which we measure \GAR. 
Because we are specifically tracking particles that at the time of interest are star forming (those that have SFR $>0$), but were not in the galaxy in the previous timestep, it is natural to expect our measure to be biased towards gas accretion that leads to star formation and hence that leads to more important changes in the RMP in the centre (where most star formation takes place\footnote{
We remind the reader that this is done for ease of comparison with observational measurements, which typically use nebular emission lines that trace HII regions.
}). 
With this technique, we are therefore likely seeing the most effect the gas accretion can have on the internal RMP. 
However, our measurement of gas accretion is a lower limit as we are by construction ignoring the gas particles that join the ISM but are not related to star formation, as well as particles that were ejected in between snapshots.

Another important caveat to keep in mind is that we are radially averaging the effect of gas accretion, which simulations show is not necessarily axisymmetric. 
This may be washing out some of the more localised effects the gas accretion can have, particularly if it is strongly filamentary (e.g., \citealt{Putman2017}, \citealt{Kacprzak2017}, \citealt{Ho2019}). 
In the future, we will study the gas accretion effects in the 2-dimensional gas metallicity distribution of galaxies with the aim of quantifying how much variation is expected in individual galaxies due to the generally asymmetric nature of gas accretion.

Despite these limitations, our work clearly establishes the important role of gas accretion in shaping the RMP of galaxies, which we are able to control for other effects, such as stellar mass and SFR variations, thanks to the statistics of the {\sc eagle} simulations, finding that {\it gas accretion is the primary responsible for the slope of the RMP}. 
Note that metallicities, SFRs, and gas fractions were not used in the process of parameter tuning in {\sc eagle} and hence, the result we present in this paper represents a true prediction of the simulation.

Our results establish some clear correlations that we would expect at fixed stellar mass: galaxies with steeper (more negative $\alpha$) RMPs at $r<r_{\rm 50}$ tend to have higher neutral gas fractions and \GAR\ (the latter being more difficult to test).

Deep measurements of the HI content of galaxies that push down to low column densities, typical of the circumgalactic medium \citep{Popping2009}, together with metallicity gradient estimates from IFS surveys, will be an ideal combination to test our predictions. 
Instruments such as the Australian Square Kilometre Array (SKA) Pathfinder (ASKAP; \citealt{Johnston2008}) and the Karoo Array Telescope (MeerKAT; \citealt{Booth2009}), and in the future the SKA, will allow the former measurements, while instruments such as SAMI \citep{Bryant2015}, MUSE \citep{Carton2018}, and other IFS surveys, allow for the latter. 
Note that our results cannot be easily extrapolated to $z>1$ as galaxy mergers are expected to become more common (as shown by \citealt{Qu2017,Lagos2018b} for {\sc eagle}) and because our \GAR\ ignores the gas that comes from galaxy mergers (as we are aiming to quantify ``smooth'' accretion), this may become an important shortfall. 
In addition, the \GAR\ at $z \gtrsim 1$ is expected to be much more collimated and to penetrate down to the galaxy more easily than at $z \lesssim 1$ (e.g. \citealt{Correa2018}). 
This may have the effect of directly feeding the central regions of the galaxy rather than from outside-in, possibly driving the inverted RMP seen in observations at $z \approx 3-4$ \citep{Troncoso2014,Cresci2010}.

Recently, \citet{Patricio2019} presented observational measurements of the RMP of $3$ strongly lensed galaxies at $z=0.6, \, 0.8$ and $1$, and reported these to have more negative $\alpha$ than $z=0$ galaxies. 
This is consistent with our findings as galaxies at $z \approx 1$ have higher \GAR, therefore leading to more negative $\alpha$. 
However, to confirm this observationally, a larger sample of galaxies is required to study the RMP at fixed stellar mass across cosmic time.

\section{Conclusions}
\label{sec:Conclusions}
In this work we use two simulations from the {\sc eagle} project, the reference large-volume, standard-resolution simulation of $100 \, {\rm Mpc}$ on a side (\simref), and the recalibrated high-resolution simulation of $25 \, {\rm Mpc}$ of a side (\simrec), to study how the slope of the radial metallicity profile (RMP) of galaxies, $\alpha$, changes with the gas accretion rate (\GAR) and other galaxy properties.
Because of the resolution limits of these simulations, we focus on galaxies with $M_\star \geq 10^{10} \, {\rm M_\odot}$ for the \simref\ simulation, and $M_\star \geq 10^9 \, \rm M_{\odot}$ for the \simrec\ simulation. 
We also limit our study to central, star forming galaxies and to redshifts $z \leq 1$.

We use a particle tracking method to find the gas particles that are being accreted onto galaxies, including those that are converted into stars. 
We are especially interested in smooth accretion rather than accretion in the form of galaxy mergers, and hence we select only those gas particles that were not part of another galaxy in the previous simulation snapshots. 
Our aim is to find whether the \GAR\ can be robustly connected to changes in the RMP as a primary driver, and hence controlling for differences in stellar mass and SFR is essential. 
The {\sc eagle} simulations are an ideal tool for this purpose, as its combination of volume and resolution allows us to look into the internal structure of galaxies as well as providing us with enough statistics to explore galaxy properties at fixed stellar mass. 
Here, we focused only on central, star forming galaxies with at least $10$ accreted gas particles in the last $\approx 100$~Myr that come from sources other than galaxy mergers, at redshifts $z=0$ and $1$.

We summarise our conclusions as follows:
\begin{itemize}

    \item 
    The gas accretion rate is positively correlated with stellar mass (Fig.~\ref{fig:MZR}) and SFR (Fig.~\ref{fig:Maccreted_SFR}), at all redshifts studied. 
    We also find that, at fixed stellar mass, higher gas metallicity galaxies are associated with lower \GAR\ (Fig.~\ref{fig:MZR}). 
    Together, these results are consistent with the MZR-SFR relation (also called fundamental metallicity relation in the literature; e.g. \citealt{Mannucci2010}).

    \item 
    A tight negative correlation is found between the inner RMP slopes (measured within $r/r_{\rm 50} \leq 1$) and \GAR\ at all redshifts studied (Fig.~\ref{fig:Slopes_Maccreted}).
    Galaxies with higher gas accretion rates tend to have steeper RMPs. 
    Even though galaxies change their \GAR\ with time at fixed stellar mass, this anti-correlation does not seem to evolve. 
    At large radii, $r/r_{\rm 50} > 1$, we find a weak trend for the RMPs of galaxies to become flatter as \GAR\ increases. 
    However, this trend is characterised by a very large scatter caused by noise, which prevents us from drawing any strong conclusion. 
    A higher resolution simulation would be required to confirm this trend.

    \item
    A clear anti-correlation between the inner RMP slope ($r/r_{\rm 50} \leq 1$) and \GAR\ remains even when eliminating the dependence on stellar mass (Fig.~\ref{fig:Residues_Maccr}). 

    \item 
    The SFR is not as strongly anti-correlated with the slope $\alpha$ as \GAR\ is (Fig.~\ref{fig:Slopes_SFR}), indicating that the latter is more fundamental in shaping the RMP.

    \item 
    We also obtain a relation between the neutral gas fraction and the slope of the inner RMP (at $r/r_{\rm 50} \leq 1$; Fig.~\ref{fig:Slopes_Fgas}), but it is less clear than with \GAR. 
    However, the gas fraction is a useful proxy as it is more readily accessible observationally than \GAR.

    \item 
    When analysing the RMP binned by redshift and stellar mass, we see that galaxies with the lowest \GAR\ show flatter inner slopes (even cored RMPs), while galaxies with the highest \GAR\ display steeper negative slopes (Fig.~\ref{fig:RMP}).
\end{itemize}

In the future, we will aim to reveal what properties are causing the high scatter at $r/r_{\rm 50} > 1$, as well as studying galaxies in two dimensions rather than radially averaged \citep{Marino2016, Trayford2019}. 
Even though the assumptions made in the calculation of \GAR\ are simple, it is clear that the latter plays a primary role in altering the RMPs of galaxies to a degree that we hope will be testable with a combination of sensitive IFS instruments, absorption line studies and deep HI observations, e.g. from local surveys such as SAMI \citep{Bryant2015} and CALIFA \citep{Sanchez2012}, and in the near future the MUSE MAGPI\footnote{https://magpisurvey.org/} survey (MAGPI collaboration in prep.) at $z\approx 0.3$.
%

\section*{Acknowledgements}
We thank the referee for their insightful and constructive report, which helped the clarity of this paper.
We thank Yannick Bah\'e for fruitful discussions and suggestions. 
We acknowledge the Virgo Consortium for making their simulation data available. The \textsc{eagle} simulations were performed using the DiRAC-2 facility at Durham,  managed by the ICC, and the PRACE facility Curie based in France at TGCC, CEA, Bruy\`{e}resle-Ch\^{a}tel. 
FC acknowledges CONICET, Argentina, and the Australian Endeavour Scholarships and Fellowships for their supporting fellowships.
CL is funded by the Australian Research Council Centre of Excellence for All Sky Astrophysics in 3 Dimensions (ASTRO 3D), through project number CE170100013. The Cosmic Dawn Center is funded by the Danish National Research Foundation. 
SAC acknowledges funding from {\it Consejo Nacional de Investigaciones Cient\'{\i}ficas y T\'ecnicas} (CONICET, PIP-0387), {\it Agencia Nacional de Promoci\'on Cient\'ifica y Tecnol\'ogica} (ANPCyT, PICT-2013-0317), and {\it Universidad Nacional de La Plata} (11-G150), Argentina. 
EW acknowledges support by the Australian Research Council Centre of Excellence for All Sky Astrophysics in 3 Dimensions (ASTRO 3D), through project number CE170100013. 

\bibliographystyle{mnras}
\bibliography{collacchioni}


\appendix

\section{Convergence of \GAR-RMP relation}
\label{sec:App-Resolution}
\begin{figure}
    \includegraphics[trim=2mm 4mm 0mm 3mm, clip,width=1.0\columnwidth]
    {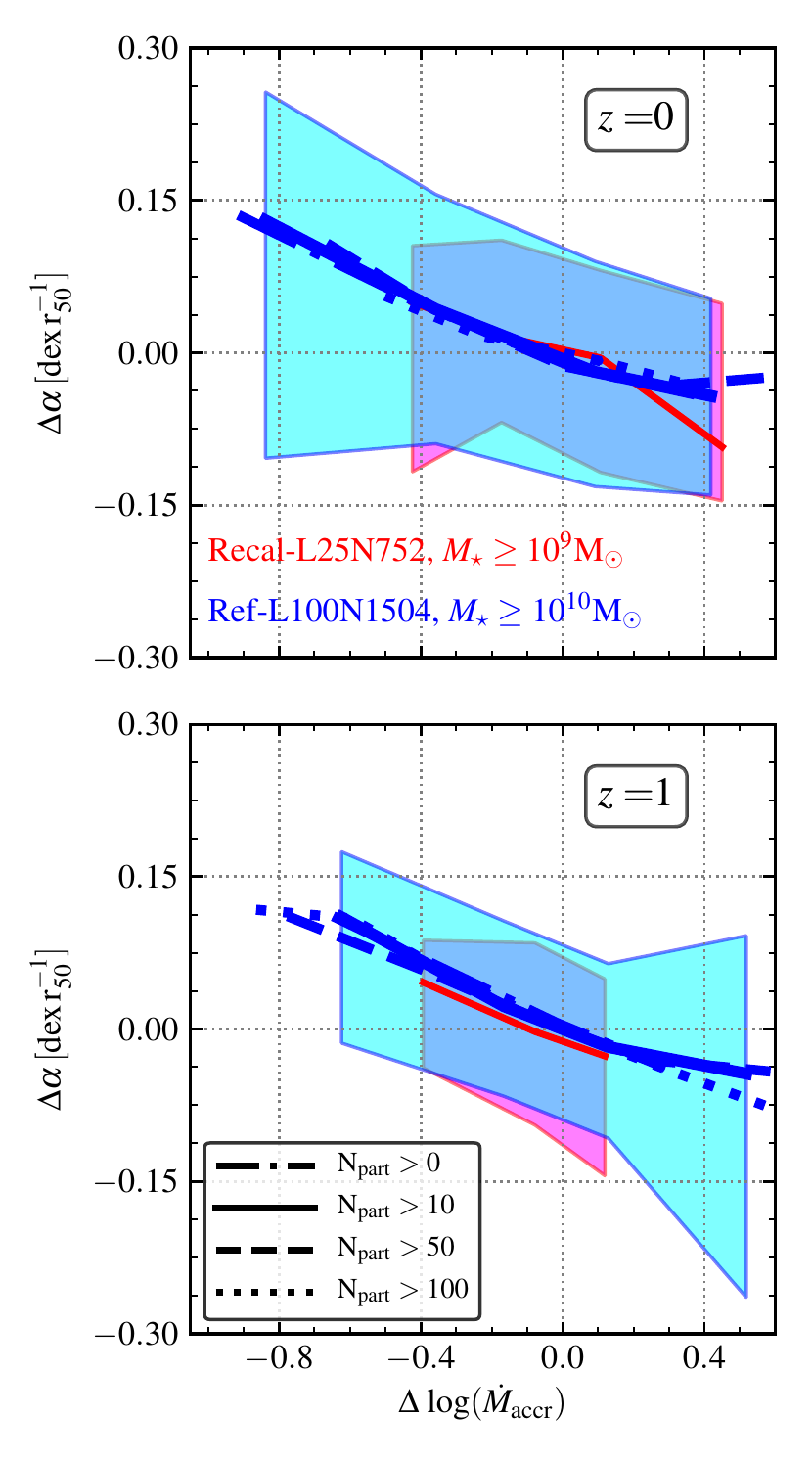}
    \caption{
    Same as Fig.~\ref{fig:Residues_Maccr}, but now comparing different minimum number of gas particles of the accreted gas above which we include galaxies in our analysis.
    Solid, dashed and dotted lines show \GAR\ in galaxies with $\ge 10$, $\ge 50$ and $\ge 100$ gas particles of accreted gas, respectively. The errorbars show the $1 \sigma$ percentiles and are shown only for the case of $\ge 10$ gas particles, for clarity.
    }
    \label{fig:Append-Residues}
\end{figure}
Throughout this work, we consider central, star-forming galaxies with at least $10$ SF gas particles that have been accreted (which were outside the galaxy $\approx 100$~Myr ago). 
This minimum of particles translates to values of \GAR\ $\approx 9 \times 10^{-2} \, {\rm M_\odot yr^{-1}}$. 
Here, we assess the robustness of our findings to the minimum number of gas accreted particles allowed. 

We show in Fig.~\ref{fig:Append-Residues} the relation between the RMP slope residuals, $\Delta \alpha$, and the \GAR\ residuals, $\Delta \dot M_{\rm accr}$, for four SF gas particles minimum number cuts, from at least $1$ particle to $100$ particles, as labelled.
We remind the reader that we define these residuals in Section~\ref{sec:Maccreted}.
For the sake of clarity, we only show the $1 \sigma$ scatter for our fiducial cut of $10$ particles for both simulations (the magenta shaded region depicts the \simrec\ simulation, while cyan represents the \simref\ simulation).

We can see that for all redshifts analysed there is almost no difference between the four SF gas particles number cuts. 
It is interesting to further analyse the cases where galaxies have fewer than $10$ accreted SF gas particles. 
For simulation \simref, 17 galaxies at $z = 0$ ($\approx 10$ per cent of the sample at this redshift) have fewer than $10$ particles and we find them to follow the same trends reported in this manuscript. 
At $z = 1$, only one galaxy from this simulation is found to have fewer than $10$ gas particles (it has \GAR\ $\approx 10^{-2.5} \, {\rm M_\odot / yr^{-1}}$ and $\alpha \approx 0.015$). 
On the contrary, simulation \simrec\ does not have any central, star-forming galaxy with fewer than $10$ accreted SF gas particles (at least until $z \leq 1$). 
We hence conclude that our results are robust to the chosen minimum number of gas particles for the accreted gas, above which we include galaxies in our analysis.

\section{Gas accretion rate as a function of stellar mass}
\label{sec:App-GARvsMstar}
\begin{figure}
    \includegraphics[trim=3mm 4mm 0mm 3mm, clip,width=1.0\columnwidth]
    {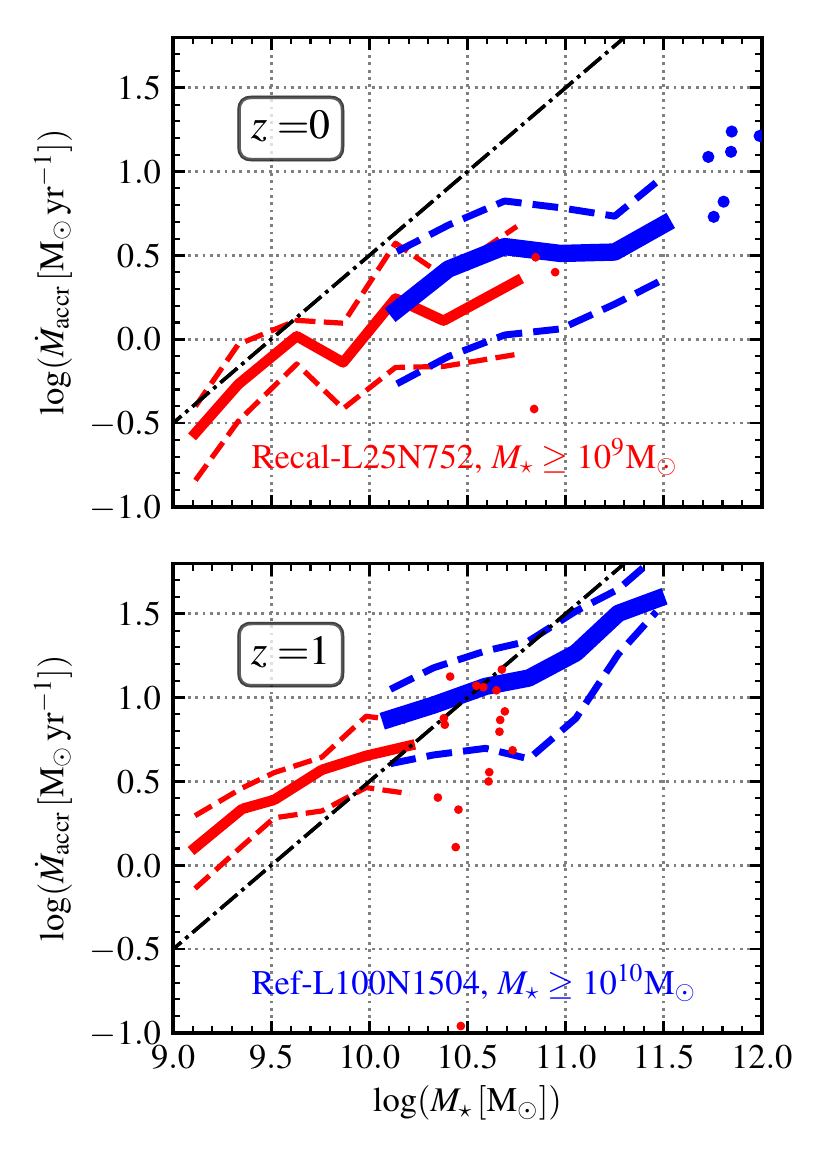}
    \caption{
    \GAR\ as a function of \mstar~at redshift $z = 0$ (top panel) and $z = 1$ (bottom panel). 
    Simulations \simrec\ and \simref\ are shown in red and blue, respectively. 
    Solid lines shown the median values of the relation, while dashed lines represent the $16^{\rm th}-84^{\rm th}$ percentile ranges. 
    Individual galaxies are shown where bins have $<10$ objects (symbols).
    The dot-dashed line shows the function $\rm log_{10}(\dot M_{\rm accr}) = log_{10}({\rm M_\star}) \, + \, 9.5$, as a reference.
    }
    \label{fig:Append-GARvsMstar}
\end{figure}
In order to understand in more detail how \GAR\ affects the chemical evolution of galaxies, it is relevant to analyse how \GAR\ depends on stellar mass, \mstar. This is the purpose of this appendix.

Fig.~\ref{fig:Append-GARvsMstar} shows the median \GAR$-$\mstar~relation at redshifts $z=0$ (top panel) and $z=1$ (bottom panel). 
Solid lines show the medians, while dashed lines represent the $16^{\rm th} - 84^{\rm th}$ percentile ranges. 
The black dotted-dashed line is a linear function with slope $\equiv  1$ to use as a reference.

As expected, the \GAR$-$\mstar~relation has a positive slope, i.e., galaxies with increasing \mstar~also show an increasing \GAR, and at fixed stellar mass, \GAR\ decreases with time.
The relation is quite tight (about $\approx 0.5$~dex of scatter, as quantified by the $16^{\rm th} - 84^{\rm th}$ percentile rang). 
A small offset between the two simulations is seen for galaxies with \mstar~$\approx 10^{10}$~\msun, though not as pronounced as the one from Fig.~\ref{fig:Slope_Mstar}. 
This offset is not unexpected due to the different environments in both simulations \citep{Furlong2015}.
Despite this, we see that once stellar mass is accounted for, a clear relation emerges between \GAR\ and the slope of the RMP.

\section{The relation between the molecular hydrogen fraction and the gas accretion rate}
\label{sec:App-GasH2}
\begin{figure*}
    \includegraphics[trim=3mm 4mm 0mm 3mm, clip,width=1.7\columnwidth]
    {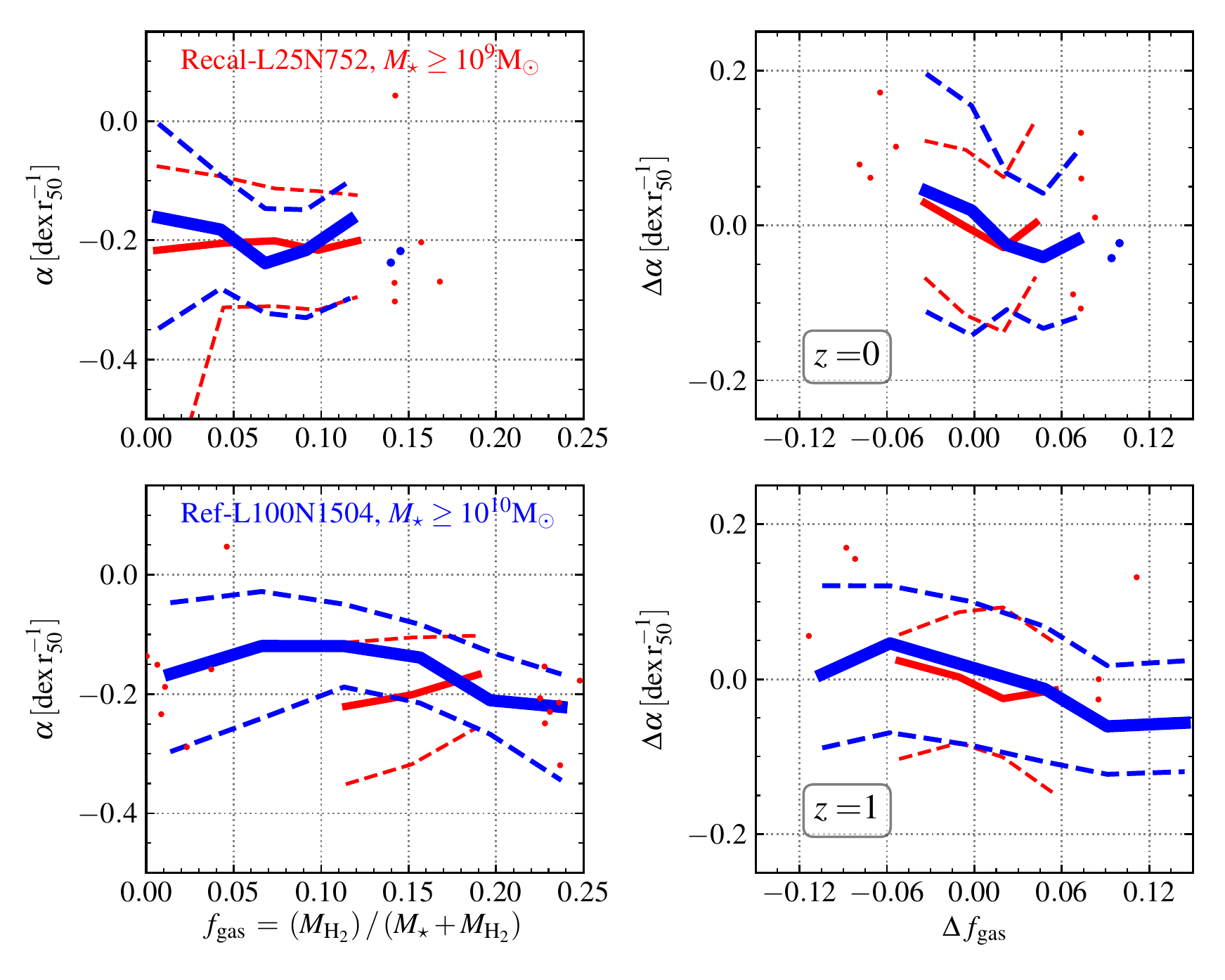}
    \caption{
    Same as Fig~\ref{fig:Slopes_Fgas}, but now taking only the contribution of the molecular gas, H$_{\rm 2}$, in the definition of the gas fraction, i.e., $f_{\rm gas} = {\rm H_2} / (M_\star + {\rm H_2})$. 
    }
    \label{fig:Append-FgasH2}
\end{figure*}
Since observationally it is easier to measure molecular gas ${\rm H_2}$ than atomic hydrogen ${\rm HI}$ at $z \gtrsim 0.5$, we test if our conclusions from Section~\ref{sec:GasFraction} might change if only considering the former. 

Fig.~\ref{fig:Append-FgasH2} is a replica of Fig.~\ref{fig:Slopes_Fgas}, but with a twist in the definition of the gas fraction. 
Now, we consider only the contribution of the ${\rm H_2}$ and so we have that gas fraction is defined as $f_{\rm gas} = M_{\rm H_2} / (M_\star + M_{\rm H_2})$. 
At first sight, one of the differences is that the values of the $f_{\rm gas}$ drop considerably compare to our previous definition. 
This is due to the fact that $M_{\rm HI}$ is more abundant than the molecular component. 
Another interesting thing to stand out is that the maximum values of $f_{\rm gas}$ increases with redshift, showing why through observations is easier to measure the $M_{\rm H_2}$ at redshift $z \gtrsim 0.5$.

With this, we conclude that whether the definition of gas fraction involves ${\rm H_2}$ or ${\rm HI}$ (not shown), the results we found in Section~\ref{sec:GasFraction} remain the same.
%

\bsp    
\label{lastpage}
\end{document}